\documentclass[paper]{JHEP3}
\pdfoutput=1
\usepackage{amsmath}
\usepackage{amssymb}
\usepackage{amsbsy}
\usepackage{bbm}
\usepackage{enumerate}
\usepackage{slashed}
\usepackage{url}
\usepackage{xspace}
\usepackage[pdftex]{graphicx}

\preprint{}

\newcommand{\muf}	{\ensuremath{\mu_F}}
\newcommand{\mur}	{\ensuremath{\mu_R}}

\newcommand{\as}	{\ensuremath{\alpha_s}}

\newcommand{\aem}	{\ensuremath{\alpha_{\rm EM}}}

\newcommand{\qq}	{\ensuremath{\mathrm{q}}}
\newcommand{\qaq}	{\ensuremath{\mathrm{\bar{q}}}}
\newcommand{\tq}	{\ensuremath{\mathrm{t}}}
\newcommand{\taq}	{\ensuremath{\mathrm{\bar{t}}}}

\newcommand{\gen}	{\ensuremath{\mathrm{gen}}}
\newcommand{\incl}	{\ensuremath{\mathrm{incl}}}
\newcommand{\isol}	{\ensuremath{\mathrm{isol}}}

\newcommand{\mt}	{\ensuremath{m_\tq}}

\newcommand\eqn[1]     {Eq.\,(\ref{#1})}

\newcommand\fig[1]     {Fig.\,\ref{#1}}
\newcommand\figs[2]    {Figs.\,\ref{#1} and~\ref{#2}}
\newcommand\figss[2]   {Figs.\,\ref{#1}--\ref{#2}}

\newcommand\tbl[1]     {Tab.\,\ref{#1}}

\newcommand\secref[1]     {Sec.\,\ref{#1}}

\newcommand{\powhel}	{\texttt{PowHel}}

\newcommand{\powhegbox}	{\texttt{POWHEG-BOX}}
\newcommand{\powheg}	{POWHEG}
\newcommand{\fastjet}	{\texttt{FastJet}}
\newcommand{\pythia}	{\texttt{PYTHIA}}

\newcommand{\pythiaver}	{\texttt{PYTHIA-6.4.25}}

\newcommand{\helacnlo}	{\texttt{HELAC-NLO}}

\newcommand{\lhapdf}	{\texttt{LHAPDF}}

\newcommand{\amcatnlo}	{\texttt{aMCatNLO}}

\newcommand{\Wgamma}	{\ensuremath{W\,\gamma}}
\newcommand{\ttgamma}	{\ensuremath{\tq\,\taq\,\gamma}}

\newcommand{\ttbar}	{\ensuremath{\tq\,\taq}}

\newcommand{\ptgamma}   {\ensuremath{p_{\bot\,,\gamma}}}

\newcommand{\pt}   	{\ensuremath{p_{\bot}}}
\newcommand{\ptj}   	{\ensuremath{p_{\bot}^j}}

\newcommand{\ptsupp}    {\ensuremath{p_{\bot\,,{\rm supp}}}}
\newcommand{\gev}	{\ensuremath{\rm GeV}}
\newcommand{\kt}	{\ensuremath{k_\bot}}

\newcommand{\ptmiss}	{\ensuremath{\slashed{p}_\bot}}

\title{Hadroproduction of t anti-t pair in association with an isolated
photon at NLO accuracy matched with parton shower}

\author{Adam Kardos\\
  MTA-DE Particle Physics Research Group,\\
  University of Debrecen, H-4010 Debrecen P.O.Box 105, Hungary\\
  E-mail: \email{kardos.adam@science.unideb.hu}
}
\author{Zolt\'an Tr\'ocs\'anyi\\ 
  Institute of Physics and MTA-DE Particle Physics Research Group,\\
  University of Debrecen, H-4010 Debrecen P.O.Box 105, Hungary\\
  E-mail: \email{zoltan.trocsanyi@cern.ch}}

\abstract{
We simulate the hadroproduction of a \ttbar-pair in association with a
hard photon at LHC using the \powhel\ package. These events are almost
fully inclusive with respect to the photon, allowing for any physically
relevant isolation of the photon. We use the generated events, stored
according to the Les-Houches event format, to make predictions for
differential distributions formally at the next-to-leading order (NLO)
accuracy and we compare these to existing predictions accurate at NLO
using the smooth isolation prescription of Frixione. Our fixed-order
predictions include the direct-photon contribution only. We also make
predictions for distributions after full parton shower and hadronization
using the standard experimental cone-isolation of the photon.
}


\begin{document}

\section{Introduction}

Isolated hard photons are important experimental tools for a variety of
processes at the LHC. Most notably, one of the cleanest channels to
identify the Standard Model (SM) Higgs particle is its decay into a pair of
hard photons. Although this channel has a small (about 0.2\,\%)
branching ratio as compared to the hadronic and leptonic channels, the
spectacular resolution of the electromagnetic calorimeters of the ATLAS
and CMS detectors and the relatively low background made this as one of
the prime discovery channels \cite{Aad:2012tfa,Chatrchyan:2012ufa}. 

From the theoretical point of view isolated hard photons are rather
cumbersome objects. Unlike leptons, the photons couple directly to
quarks. If the quark that emits the photon is a light quark, treated
massless in perturbative QCD, then the emission is enhanced at small
angles and in fact, becomes singular for strictly collinear emission.
The usual experimental definition of an isolated photon allows for
small hadronic activity even inside the isolation cone. Due to the
divergence of the collinear emission, this isolation cannot be
implemented directly in a perturbative computation at leading-order
(LO) accuracy because even small hadronic activity inside the cone
leads to infinite results.  

Of course, one can approximate the experimental definition with complete
isolation of the photon from the coloured particles inside a fixed cone
and obtain a perturbative prediction at LO. The problem however, comes
back with a different face if we want to define the isolated photon in a
computation at the next-to-leading order (NLO) accuracy. At NLO there are
two kinds of radiative corrections: (i) the virtual one with the same
final state as the Born contribution, but including a loop and (ii) the
real one that involves the emission of a real parton in the final state.
These two contributions are separately divergent, but their sum is finite
for infrared (IR) safe observables according to the KLN theorem
\cite{Kinoshita:1962ur,Lee:1964is}.
The IR-safe observables are represented by a jet function $J_m$, where
$m$ is the number of partons in the final state: for an $n$-jet measure
$m=n$ at LO and for the virtual corrections, while $m=n+1$ in the real
correction.

There exist general methods (see e.g.~ref.~\cite{Catani:2002hc}) to
combine the real and virtual corrections for infrared (IR) safe
observables $J_m$, for which $J_{n+1}$ tends to $J_n$ smoothly in
kinematically degenerate regions of the phase space, namely when two
final-state partons become collinear or a final-state gluon becomes
soft. The problem with the isolated-photon cross section in
perturbative QCD is that the cone-photon isolation is not IR safe
beyond LO. The extra gluon in the real radiation may be radiated within
the isolation cone in which case the event will be cut even if the
gluon energy tends to zero.  

There are ways to make predictions for photon production in perturbation
theory, but all have drawbacks. In a pioneering work
\cite{Kunszt:1992np} the measurement of the inclusive photon cross
section was advocated, but that is not very useful from the
experimental point of view. In ref.~\cite{Frixione:1998jh} an isolation
procedure was proposed that is similar in spirit to the case of
inclusive cross section, yet provides a smooth isolation prescription
that is IR safe at all orders in perturbation theory. However, the
implementation of the smooth prescription experimentally is very
cumbersome as it requires very fine granularity of the detector, so it
has never become popular among experimenters.

There is a precise way of defining the isolated photon theoretically, but
that requires the inclusion of the photon fragmentation component as
well (see e.g.~\cite{Catani:1998yh}). The drawback of this approach is
the need for non-perturbative input and the extra computational effort
for a contribution that is mostly discarded when the experimental
isolation is used (cone with small hadronic activity inside that is
described by the fragmentation contribution). Thus one would be tempted
to neglect the fragmentation contribution, which is however,
uncontrolled from the theoretical point of view and thus is not a
viable option in a fixed-order computation.

In the last decade new approaches were proposed to make predictions that
are formally accurate to NLO but include the advantage of event
simulations of the shower Monte Carlo (SMC) programs
\cite{Frixione:2002ik,Nason:2004rx,Frixione:2007vw}. By now many
processes have been included in the generic frameworks of these NLO+PS
approaches, the \amcatnlo\ \cite{Alwall:2014hca} and the \powhegbox\
\cite{Alioli:2010xd} codes. In a series of papers we combined the
\powhegbox\ with the \helacnlo\ package \cite{Bevilacqua:2011xh} into
\powhel\ to make predictions for the hadroproduction of a \ttbar-pair
in association with a hard boson (scalar \cite{Garzelli:2011vp},
pseudoscalar \cite{Dittmaier:2011ti}, vector \cite{Garzelli:2012bn}
or jet \cite{Kardos:2011qa}). The only missing boson of the SM in this
list is the hard photon. In view of the above, the reason is clear: the
photon has to be isolated, which makes this computation more involved
than for the other cases.

In this paper we use the \powhel\ framework to simulate events
containing direct photons only, that is we neglect the fragmentation
contribution.  We generate the events with loose isolation cut,
resulting in an almost inclusive event sample. We argue that with
sufficiently loose generation isolation the fragmentation contribution
should be indeed small. The output of the \powhegbox\ consists of
pre-showered events stored according to the Les Houches accord (LHEs)
\cite{Alwall:2006yp}.  The LHEs when fed into a SMC, result in showered
events on which the usual experimental cone isolation can be applied.
We discuss the validity of this approach on the example of \Wgamma\
hadroproduction for which predictions including a modelling of
fragmentation as well as experimental results exist. Using events
generated with loose isolation cuts, we make predictions to \ttgamma\
hadroproduction, but the approach is general and can be used to make
predictions for any other process that involves isolated hard photons
in the final state at NLO accuracy matched with PS.  

\section{Details of the implementation}

\powhel\ is a computational framework composed of the \powhegbox\ 
\cite{Alioli:2010xd} and the \helacnlo\ \cite{Bevilacqua:2011xh}
packages to provide predictions at the hadron level with NLO QCD
accuracy in the hard process.  The essential ingredients needed for a
particular process are the matrix elements for the Born, virtual and
real-emission contributions, spin- and colour-correlated matrix elements
and a suitable phase space for the Born process. 

The matrix elements are provided by \helacnlo\ while the Born phase
space is constructed by us using the relatively simple kinematics at
the Born level. The Born phase space is generated with the help of one
kinematic invariant and three angles. An overall azimuth is
kept fixed and randomly reinstated at the end of the calculation as a
common practice in \powhegbox. Matrix elements are generated for the
following subprocesses: $\qq\,\qaq\to\gamma\,\tq\,\taq$,
$g\,g\to\gamma\,\tq\,\taq$ (tree-level for the Born process and at
one-loop for the virtual) and $\qq\,\qaq\to\gamma\,\tq\,\taq\,g$, 
$g\,g\to\gamma\,\tq\,\taq\,g$ for the real emission
$(\qq\in\{{\rm u,d,c,s,b}\})$.  The ordering among particles
follows the convention of \powhegbox: non-QCD particles, massive
quarks, massless partons. Matrix elements for all other
subprocesses are obtained from these by means of crossing.

All matrix elements, including the crossed ones, are compared to the
stand-alone version of \helacnlo\ in several, randomly chosen
phase-space points. The internal consistency between the Born, spin-,
colour-correlated and real-emission matrix elements is checked by
comparing the limit of the real-emission part and the corresponding
counter terms in all kinematically degenerate regions of the phase
space.

\begin{figure}
\includegraphics[width=0.50\textwidth]{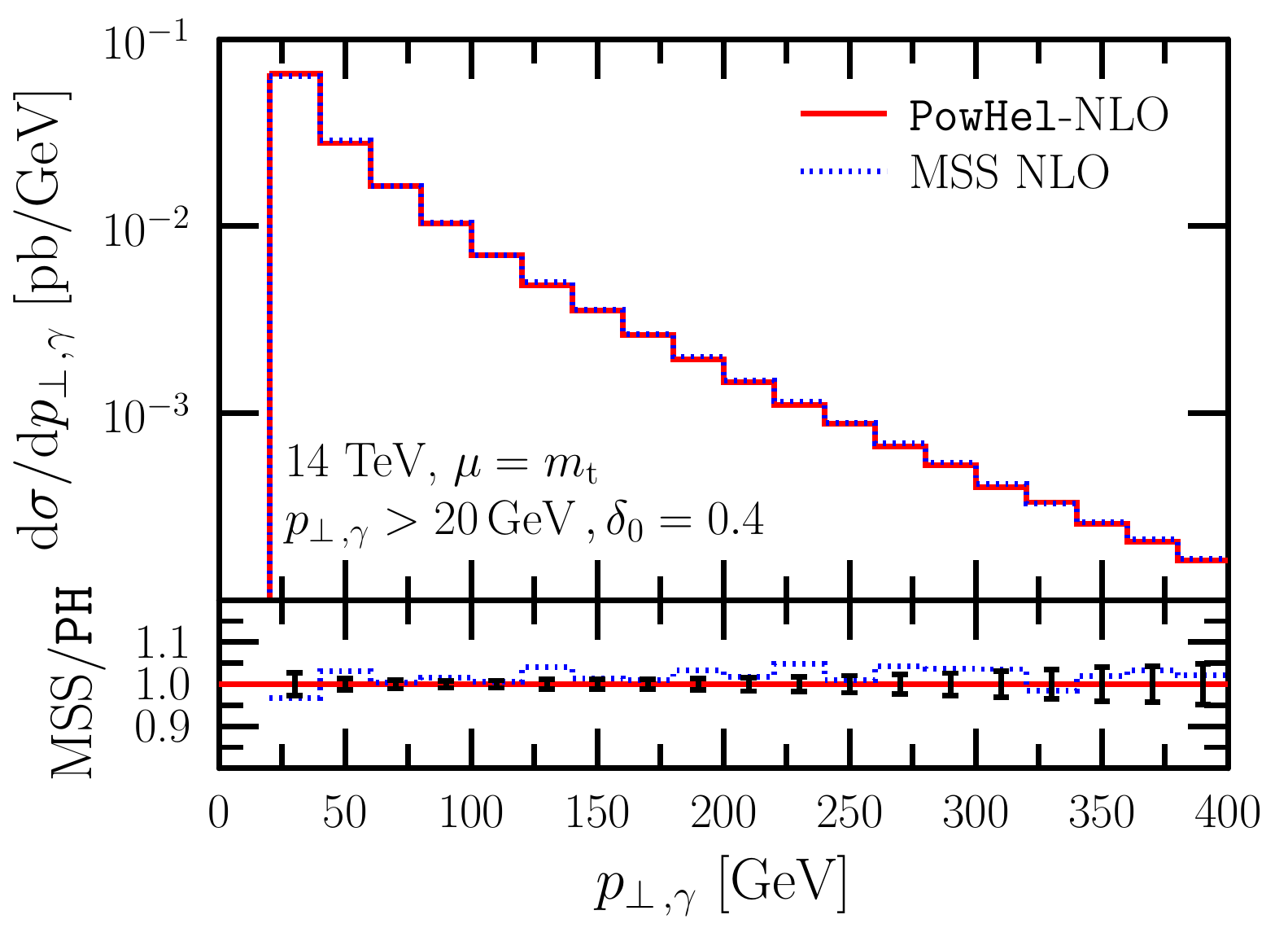}
\includegraphics[width=0.50\textwidth]{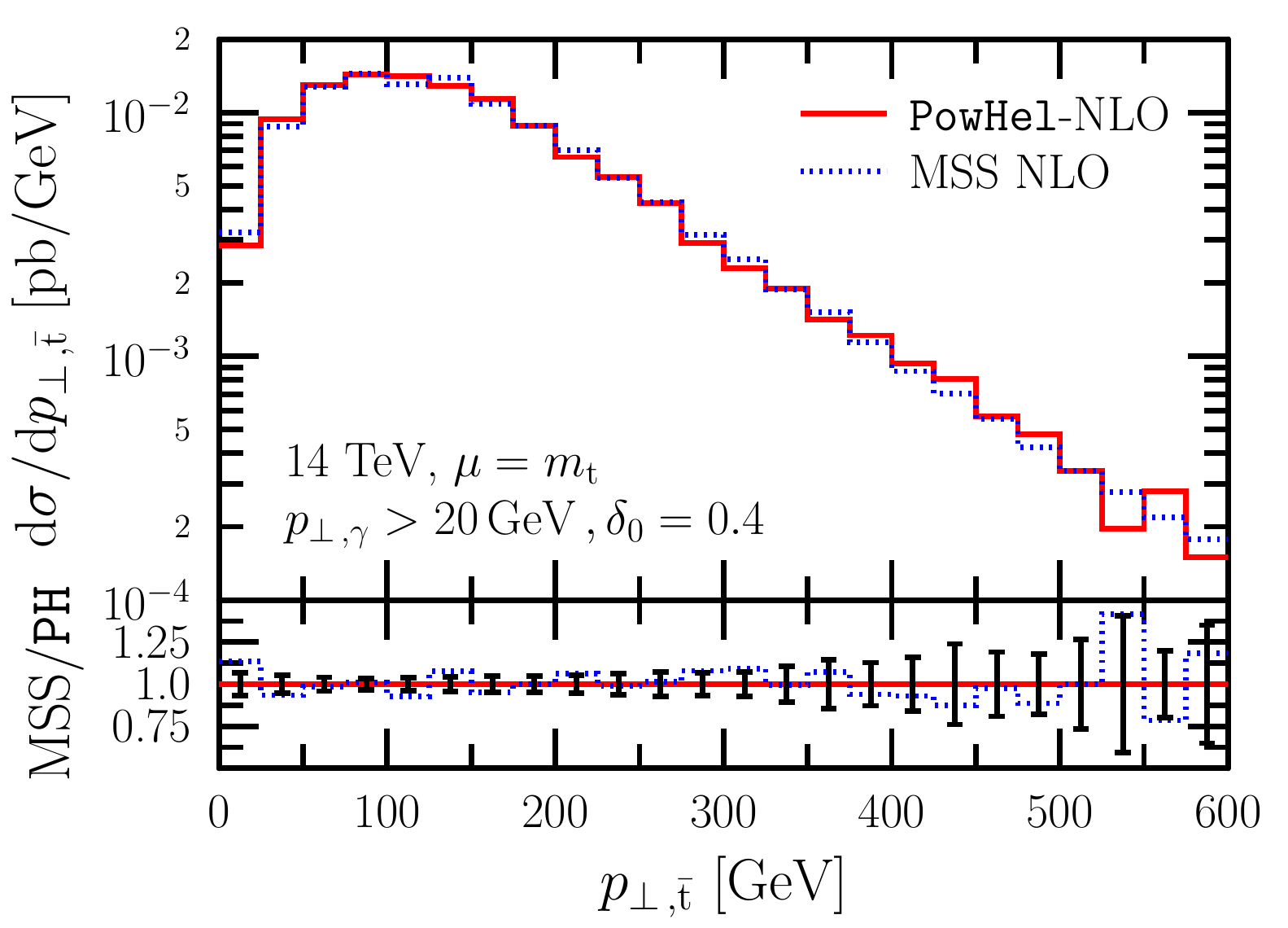}
\caption{\label{fig:MSScomp} Comparison between \powhel\ and 
\cite{Melnikov:2011ta} at the central scale with NLO accuracy for
the differential cross section as a function of the transverse
momentum of the photon and anti-t quark. Lower panels depict the
ratio of predictions in \cite{Melnikov:2011ta} (MSS) to ours. The
uncertainties appearing on the lower panels only take into account the
statistical uncertainty of our calculation.}
\end{figure}

In order to check the whole implementation we compare differential
distributions to those in ref.~\cite{Melnikov:2011ta} using the LHC
setup in the published paper: the calculation was performed for LHC at
centre-of-mass energy$\sqrt{s} = 14$\,TeV with a \texttt{CTEQ6L1} and
\texttt{CTEQ6.6M} PDF at LO and NLO accuracy and a one- and two-loop
running \as, respectively.  The mass of the t-quark was $\mt =
172\,\gev$, the fine-structure constant, was set to $\aem = 1/137$. The
renormalization and factorization scales were set fixed, equal to
$\mt$.  In the analysis a photon was required to be hard,
$\ptgamma\! >\! 20\,\gev$ and the smooth isolation of Frixione
\cite{Frixione:1998jh} was employed with isolation parameters
$\delta_0 = 0.4$ and $\epsilon_\gamma = n = 1$.
The cross sections obtained with \powhel\ are enlisted on
\tbl{tbl:MSSxscheck}.  We found complete agreement with the predictions
of \cite{Melnikov:2011ta} both for the cross sections and for the
available distributions as well. Two out of these are depicted in
\fig{fig:MSScomp}.  
\begin{table}
\begin{center}
\begin{tabular}{|c|c|c|}
\hline
\hline
$\mu$   & $\sigma_{\rm LO}^{\rm PH}$ $[{\rm pb}]$ & 
$\sigma_{\rm NLO}^{\rm PH}$ $[{\rm pb}]$ \\
\hline
$2\mt$  & $1.519 \pm 0.004$ & $2.58\pm0.03$ \\
\hline
$\mt$   & $1.968 \pm 0.005$ & $2.94\pm0.04$ \\
\hline
$\mt/2$ & $2.614 \pm 0.006$ & $3.33\pm0.06$ \\
\hline
\hline
\end{tabular}
\end{center}
\caption{{\label{tbl:MSSxscheck}} Cross sections obtained with \powhel\
at LO and NLO accuracy using the setup and cuts of \cite{Melnikov:2011ta}.
The renormalization and factorization scales are made equal to $\mu$.}
\end{table}

Having checked the implementation of the NLO computation, we generated
events with the \powhegbox.  The final state in the Born contribution,
\ttgamma, is composed of two massive and one massless particles. The
cross section when the photon is emitted from a massless (anti)quark
can become singular. This can happen when the photon is emitted by one
massless (anti)quark from the initial state, or from a final state one
in the real-emission contribution. These configurations have to be
avoided such that the physical cross sections for isolated photon
production do not depend on the actual implementation.

Let us first focus only on the singular radiation present at the Born
level. In this case there are two simple solutions to avoid
infinite contributions to the cross section. The first is a
generation cut \cite{Kardos:2011qa}, which if applied on the transverse
momentum of the photon, can avoid the singularity. This cut has to be
sufficiently small so that when physical cuts are applied, the
prediction becomes independent of this generation cut. Although this
method offers an easy way to avoid the singularity, yet we end up
generating events mostly with photons having small transverse momentum.
Hence the majority of events will be generated in a region of phase
space which has no physical importance and the efficiency of the event
generation is small.  

The other solution is the inclusion of a suppression factor
\cite{Alioli:2010qp} which can be used to enhance event generation in
certain regions of the phase space. The distributions are always
independent of the suppression used as events are generated according
to a cross section obtained with a suppression factor, but the weight of
each event is multiplied by the inverse of the suppression factor (for
details see \cite{Kardos:2014dua}). In our calculation the analytical
form of suppression was chosen to be
\begin{align}
\mathcal{F}_{\rm supp} &= 
\frac{1}{1 + \left(\frac{\ptsupp^2}{\ptgamma^2}\right)^i}
\label{eqn:suppfact}
\,,
\end{align}
and we found $i=2$ a suitable choice and $\ptsupp = 100\,\gev$ was set
throughout the whole calculation. 
 It is not necessary, yet we included
also the generation cut on the transverse momentum of the photon, by
requiring the transverse momentum of the photon in the underlying phase
space to be larger than 15\,GeV. We checked that this cut does not affect
our predictions with physical cuts larger than 15\,GeV. Our strategy to
handle singularities coming from collinear photon-emission from final
state massless (anti)quarks will be covered in the next section.

In order to speed up the event generation the real emission part can
be decomposed into a singular and finite contribution such that the
former contains all the kinematically degenerate regions of phase space,
while the latter is finite over the whole phase space. When this
decomposition is implemented the \powheg\ Sudakov factor is evaluated
with only the singular contribution. For the decomposition we used the
original suggestion of \cite{Alioli:2010xd} which became standard in
all calculations done with the help of \powhegbox. Beside of this
decomposition and the generation isolation nothing is taken into
account which could alter the shape of the \powheg\ Sudakov, that is
the matching systematics. In particular, we have not used the
\verb+hfact+ option which is only used in \cite{Alioli:2008tz} and in
all the other cases the separation of real emission contribution,
mentioned above, was considered only.

\section{NLO-LHE comparison}
\label{sec:NLO-LHEcomp}

\begin{figure}
\begin{center}
\includegraphics[width=0.80\textwidth]{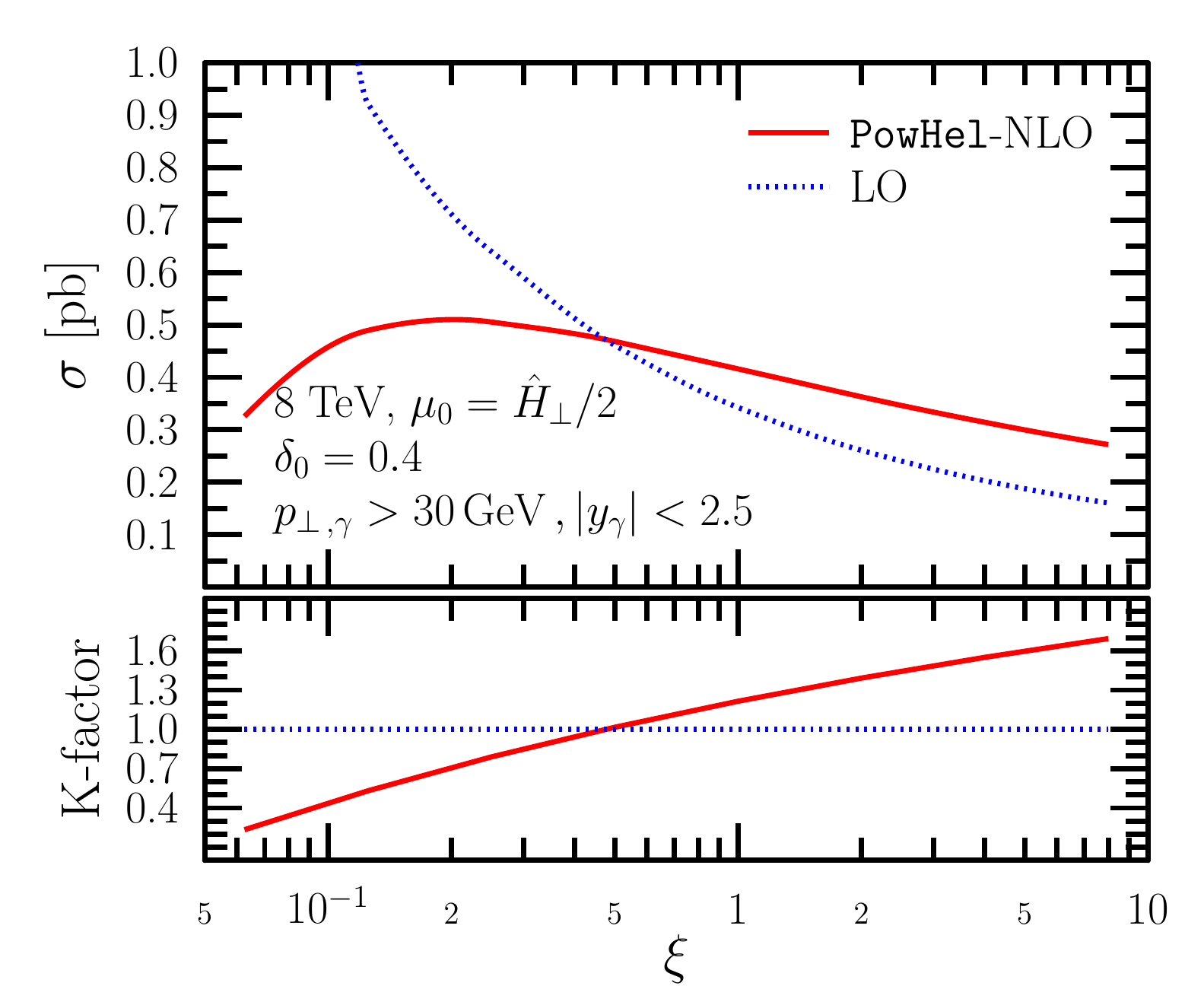}
\caption{\label{fig:Kfactor} Cross section with cuts
listed in the text and also shown in the figure at LO (blue dotted)
and at NLO (red solid) accuracy as a function of the equal
renormalization and factorization scale normalized to the default
scale $\mu_0$. The lower panel shows the NLO K-factor.}
\end{center}
\end{figure}

In this and all the upcoming sections predictions are made for
proton-proton collisions at $\sqrt{s} = 8$\,TeV with the following
parameters: \texttt{CT10nlo} PDF using \lhapdf\ \cite{Whalley:2005nh}
with a 2-loop running \as\ considering 5 massless quark-flavours, $\mt =
172.5\,\gev$, $\aem = 1/137$. For our default scale we decided to use a
dynamical one, the half of the sum of transverse masses of all
final-state particles:
\begin{align}
\mu_0 = \frac{1}{2}\hat{H}_\bot = \frac{1}{2}
\left(
m_{\bot\,,\tq} + m_{\bot\,,\taq} + \ptgamma
\right)
\,,
\end{align}
where the hat reminds us that underlying-Born kinematics was used to
evaluate the sum. For the NLO-LHE comparison the following set of
cuts was employed:
\begin{itemize}
\itemsep=-2pt
\item The photon had to be hard enough, $\ptgamma > 30\,\gev$.

\item The photon was constrained into the central region, $|y_\gamma| < 2.5$.

\item To avoid the quark-photon singularity a Frixione-isolation was used
with $\delta_0 = 0.4$ and $\epsilon_\gamma = n = 1$.
\end{itemize}
The cross section at LO and NLO accuracy as a function of the equal
renormalization and factorization scale normalized to the default
scale $\mu_0$ is shown in \fig{fig:Kfactor}. We find significant
reduction of the scale dependence and an NLO K-factor $K=1.21$
at our default scale choice.

\begin{figure}
\includegraphics[width=0.50\textwidth]{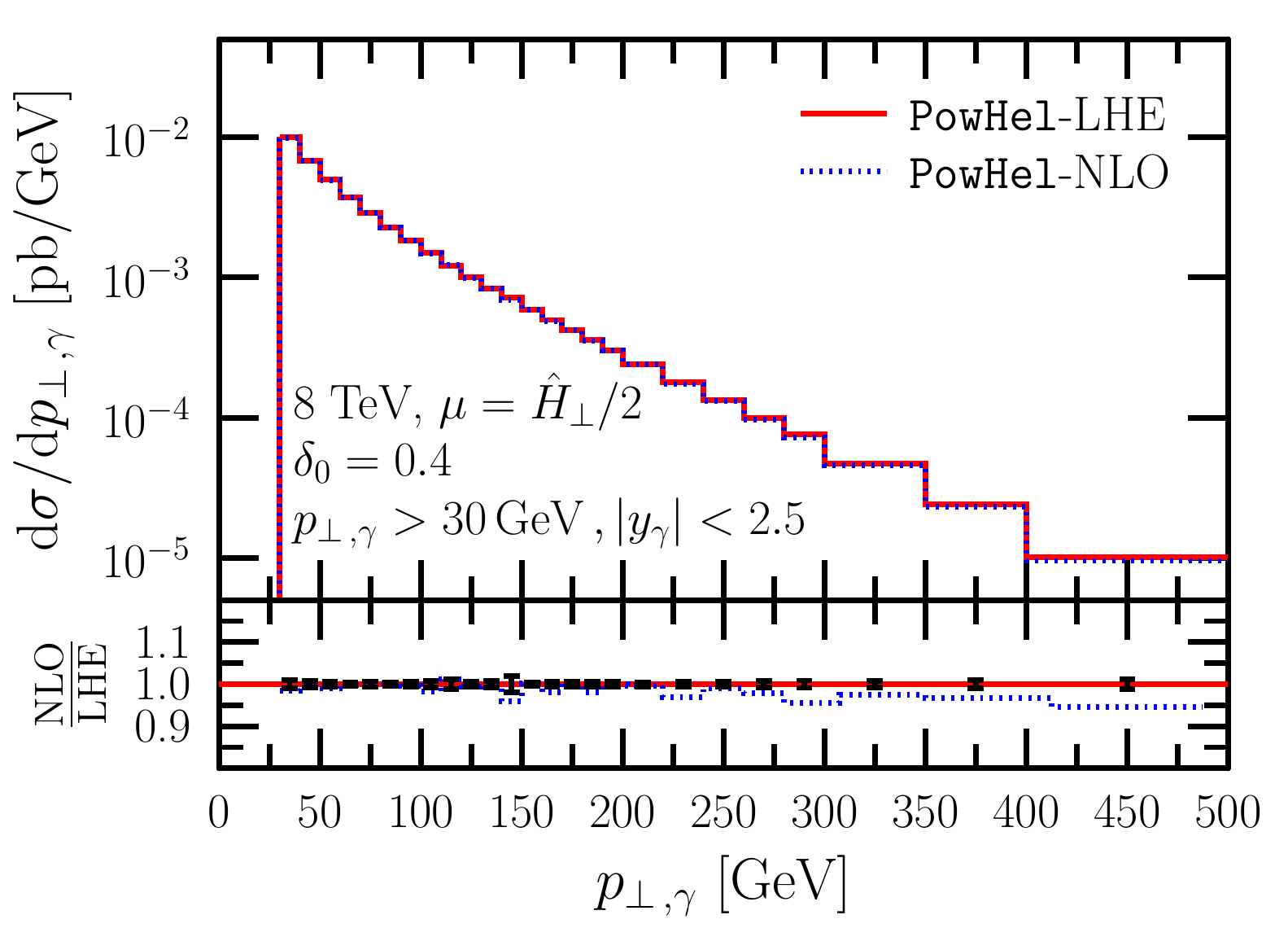}
\includegraphics[width=0.50\textwidth]{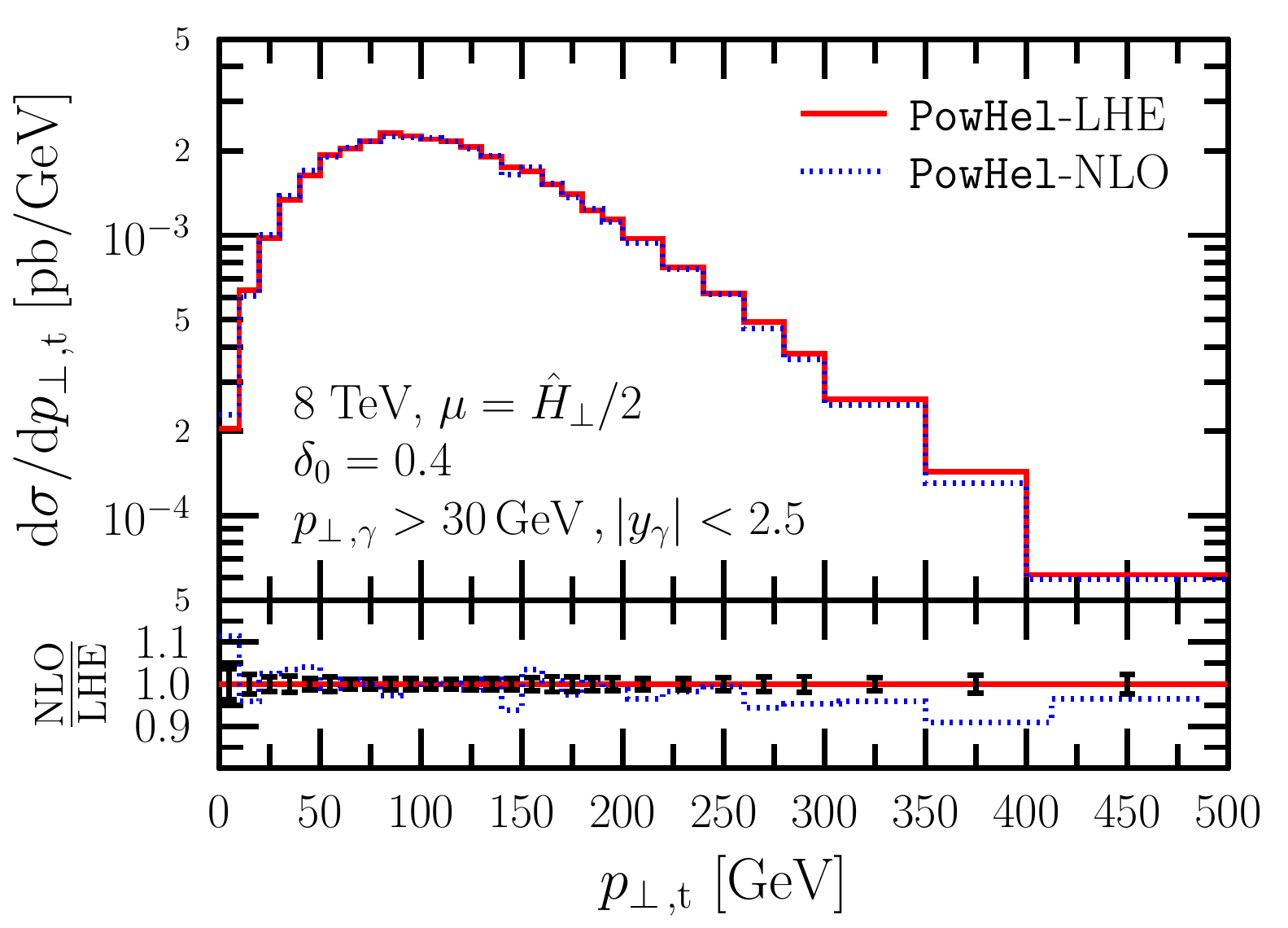}
\caption{\label{fig:LHE-NLOcompFx-pta-pttq} Comparison between
predictions from LHEs (solid red) and at NLO (blue dashed) using
Frixione isolation for the transverse momentum of the photon and
t-quark. On the lower panel the ratio of the two predictions is shown.}
\end{figure}

\begin{figure}
\includegraphics[width=0.50\textwidth]{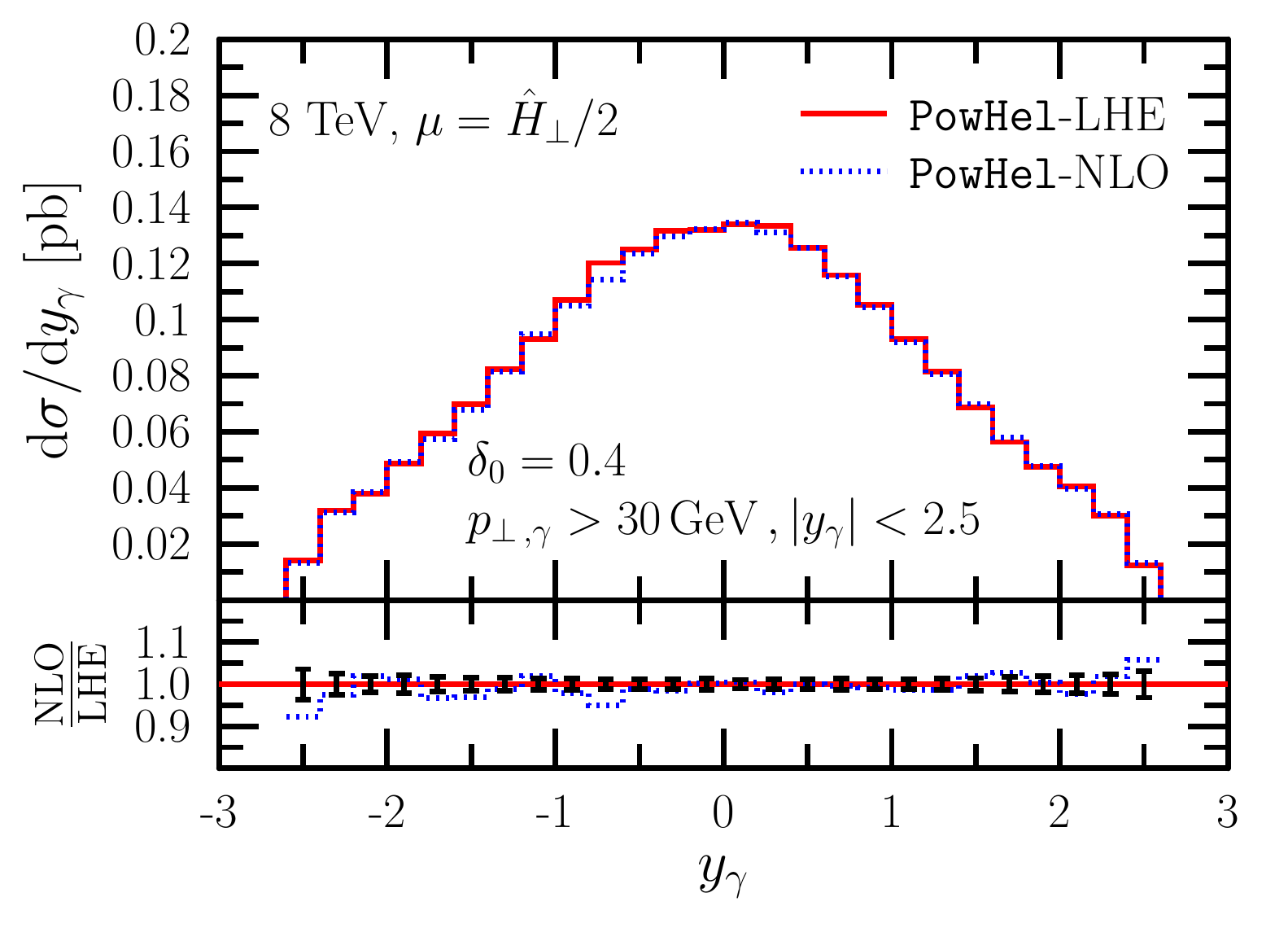}
\includegraphics[width=0.50\textwidth]{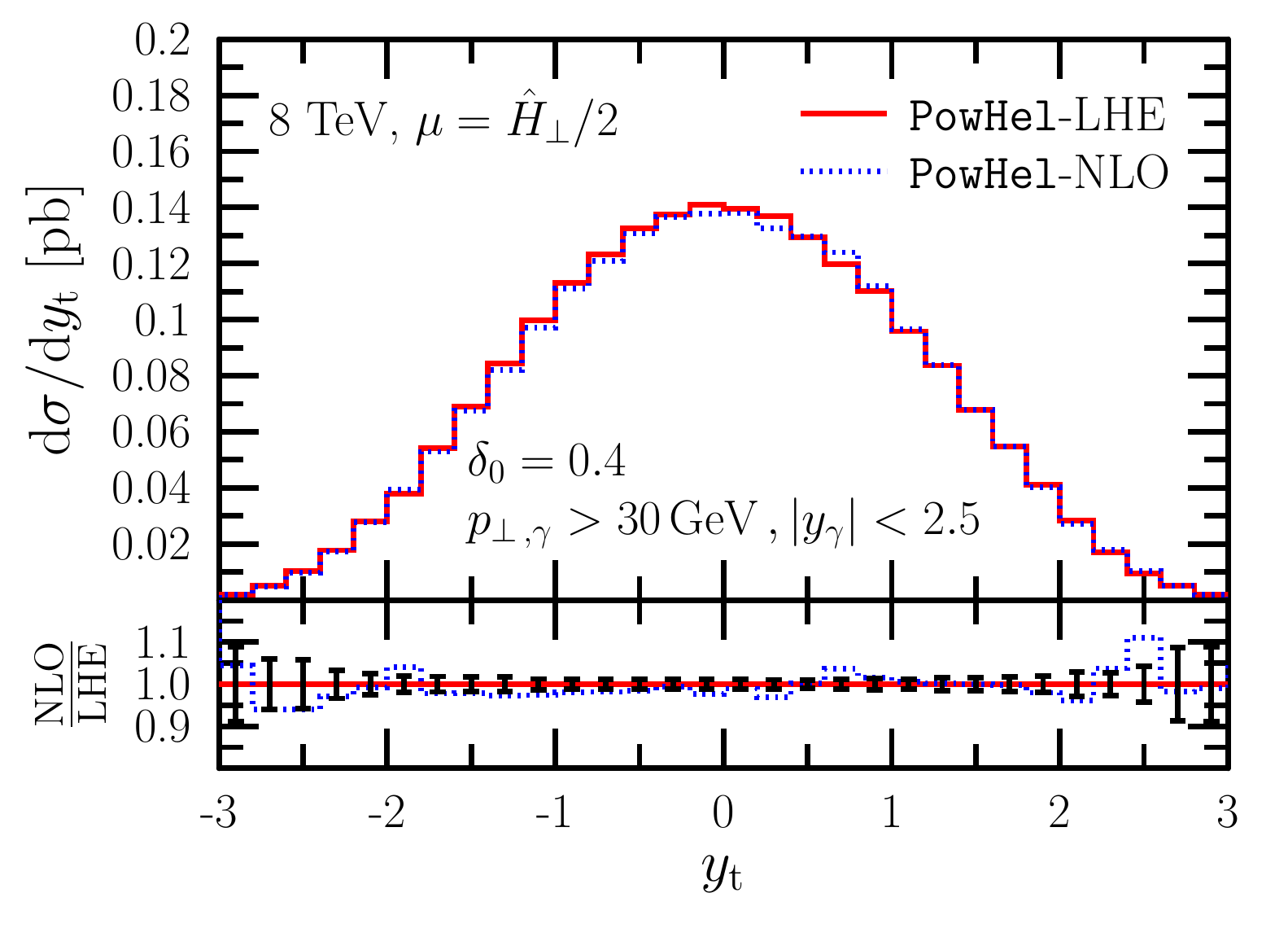}
\caption{\label{fig:LHE-NLOcompFx-ya-ytq} The same as 
\fig{fig:LHE-NLOcompFx-pta-pttq} for the rapidities of the photon and the
t-quark.}
\end{figure}

\begin{figure}
\includegraphics[width=0.50\textwidth]{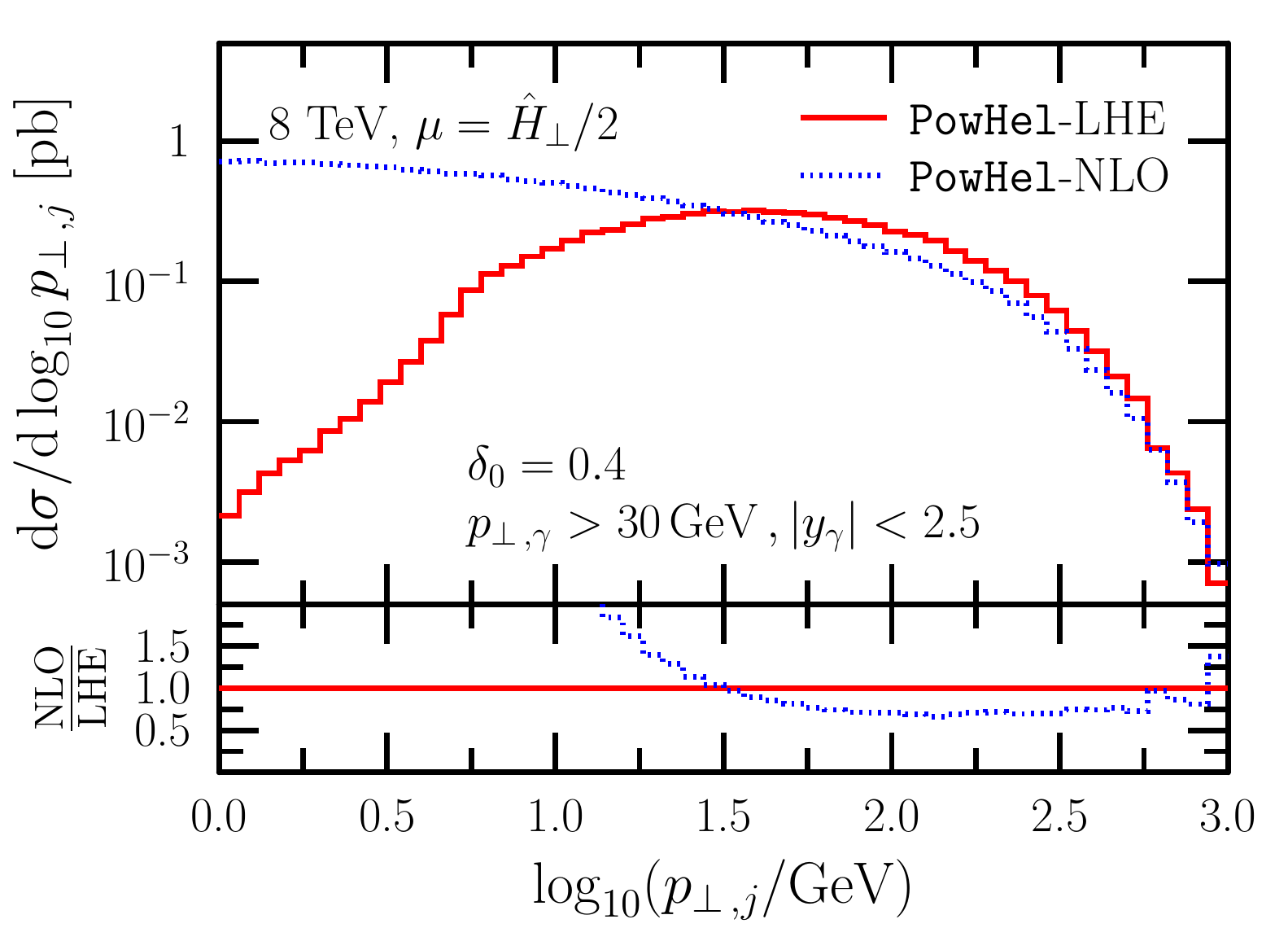}
\includegraphics[width=0.50\textwidth]{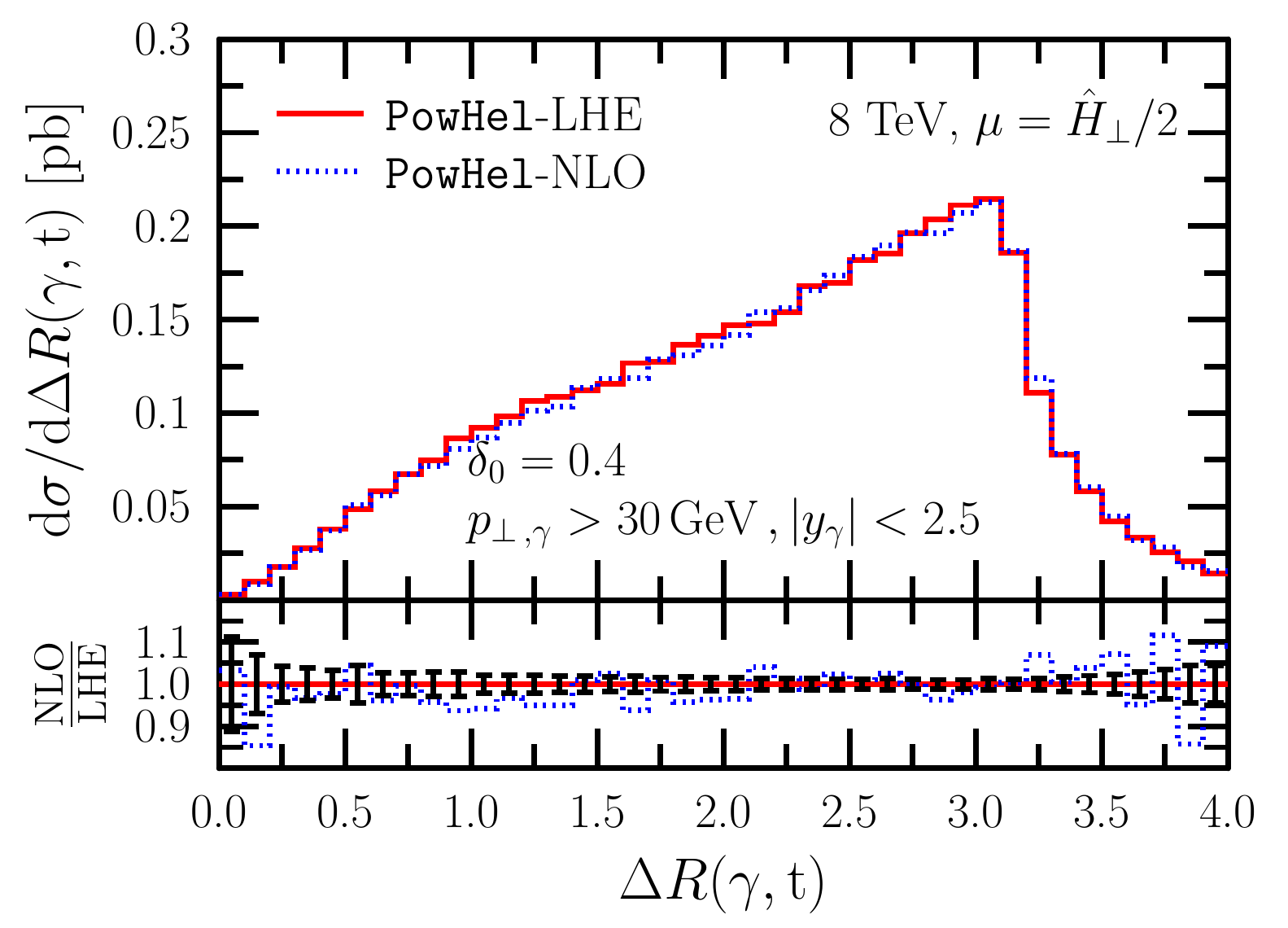}
\caption{\label{fig:LHE-NLOcompFx-log10ptj-dRatq} The same as 
\fig{fig:LHE-NLOcompFx-pta-pttq} but the differential cross section is
depicted as a function of the logarithm of the extra-parton transverse 
momentum and the separation of the photon and the t-quark. The
separation is defined in the rapidity--azimuthal angle plane.}
\end{figure}

Next we turn to comparisons of predictions at NLO accuracy with those
obtained from the pre-showered events.  With this comparison our only
aim is to demonstrate that our framework can generate meaningful
pre-showered events using the Frixione isolation (the standard in
fixed-ordered calculations).  On \figss{fig:LHE-NLOcompFx-pta-pttq}
{fig:LHE-NLOcompFx-log10ptj-dRatq} six sample distributions are
depicted to illustrate the effect of the \powheg\ Sudakov factor. In general
we find agreement between the corresponding predictions except for the
transverse-momentum distribution for the extra parton (left plot of
\fig{fig:LHE-NLOcompFx-log10ptj-dRatq}).  The effect of the \powheg\
Sudakov suppression is clearly visible in the low \pt\ region where the
radiation activity is highly limited, as expected. The presence of
the extra cut in the real-emission part (the Frixione isolation) causes
small distortion in the Sudakov shape as seen at about 0.75.
in the left plot of \fig{fig:LHE-NLOcompFx-log10ptj-dRatq}.
These comparisons show good agreement between the fixed-order
predictions and those from the pre-showered events. The visible
differences can be accounted for the effect of the \powheg\ Sudakov
factor. It is worth mentioning that the formal accuracy is still NLO,
the difference is due to higher order terms.

\section{Photon isolation as a generation cut}
\label{sec:photonisol}

When photons are produced with massless partons in the final state the
usual soft/collinear divergences coming from parton-parton splittings
are accompanied by a new type of collinear splitting, namely the
quark-photon one. The singularity produced by a collinear photon
emission off a massless (anti)quark can be absorbed into the photon
fragmentation function, decomposing the cross section into direct
photon production and a fragmentation contribution.

The only known solution that leads to an IR-safe cross section at all
orders in perturbation theory that avoids the fragmentation contribution
is offered in ref.~\cite{Frixione:1998jh} where QCD activity is
considered in a continuously shrinking cone around the photon such that
the allowed activity decreases with decreasing cone size.

While in a theoretical calculation the shrinking cone size can be
easily implemented, in an experiment the finite resolution of the
detector does not allow for taking the smooth limit.  As a result most
of the experiments adopt a different isolation criterion:  reduced
hadronic activity is allowed around the photon in a cone with finite
size such that for the total hadronic transverse energy inside the cone
\begin{align}
E_{\perp\,,{\rm had}} = \sum_{i\in{\rm tracks}}E_{\perp\,,i} \Theta
\left(
R_\gamma - R(p_\gamma,p_i)
\right)
< E_{\perp\,,{\rm had}}^{\rm max}
\,.
\label{eqn:expisol}
\end{align}
In \eqn{eqn:expisol} $E_{\perp\,,i}$ is the transverse energy of the
$i$th track, $R_\gamma$ is the isolation cone size, $R(p_\gamma,p_i)$
is the separation between the photon and the $i$th track measured in
rapidity--azimuthal angle plane, while $E_{\rm had}^{\rm max}$ is the maximal
hadronic energy allowed to be deposited in the cone of $R_\gamma$
around the photon. In the following we call this quantity hadronic or
partonic leakage depending on whether the process is considered on the
hadron or the parton level. In a fixed-order calculation an isolation
of the form of \eqn{eqn:expisol} does not completely remove the singularity of
collinear quark-photon emission and therefore, cannot be applied.
Setting $E_{\perp\,,{\rm had}}^{\rm max} =0$ removes this singularity,
but cuts into the phase space of soft gluon emission in the real
correction, hence it is not IR-safe. 


As shown in the previous section, there is one photon isolation which
is free from perturbative singularities and can be used to generate
meaningful pre-showered events.  
When generating LHEs according to the POWHEG formula, we can generate
events using the smooth isolation prescription of the photons according
to the formula (Frixione-type isolation with $\epsilon_\gamma = n = 1$)
\begin{equation}
E_{\perp\,,{\rm had}} = \sum_{i\in{\rm partons}}E_{\perp\,,i} \Theta
\left( \delta - R(p_\gamma,p_i) \right) \leq E_{\perp\,,\gamma}
\left(\frac{1-\cos\delta}{1-\cos\delta_0}\right)
\,,
\label{eq:smooth}
\end{equation}
for all $\delta \leq \delta_0$, where $\delta_0$ is a sufficiently
small, pre-defined number. This isolation can be considered as a
generation isolation, $\Theta_\isol^\gen(\delta_0)$. Then the
inclusive cross section can be decomposed as
\begin{equation}
\sigma_\incl = \sigma_\incl\,\Theta_\isol^\gen
+ \sigma_\incl\,\big(1-\Theta_\isol^\gen\big)
\,,
\label{eq:sigmahard}
\end{equation}
where the first term on the right hand side is the perturbatively 
computable cross section with smooth photon isolation of \eqn{eq:smooth}.
The second one contains non-perturbative contribution to, therefore,
cannot be computed in perturbation theory. Thus \eqn{eq:sigmahard}
can be considered as (an unconventional) factorization of the
quark-photon singularity into non-perturbative contribution.
Applying a physical isolation on \eqn{eq:sigmahard}, we obtain the 
experimentally measurable cross section for isolated photon production,
\begin{equation}
\sigma_\isol^{\rm exp} =
\sigma_\incl\,\Theta_\isol^{\rm exp} =
  \sigma_\incl\,\Theta_\isol^\gen(\delta_0)\,\Theta_\isol^{\rm exp}
+ \sigma_\incl\,\big(1-\Theta_\isol^\gen(\delta_0)\big)\,\Theta_\isol^{\rm exp}
\,.
\label{eq:sigmaisolexp}
\end{equation}
If the experimental isolation is simply a tighter version of the smooth
isolation of \eqn{eq:smooth}, then the non-perturbative contribution
trivially vanishes, as 
$\big(1-\Theta_\isol^\gen(\delta_0)\big)\,\Theta_\isol^{\rm exp} = 0$.
Thus the events generated with smooth isolation can be used to make
such physical prediction. If the physical isolation is the cone-type
isolation of \eqn{eqn:expisol}, then the non-perturbative contribution is
non-zero. Nevertheless, we shall argue that if the generation isolation
is sufficiently loose and the photon is sufficiently hard, then for
cone-type isolation with values of parameters used in the experiments,
the non-perturbative contribution is negligible within the accuracy of
the perturbative one, thus the first term still gives a sufficiently
good description of data.  

First let us note that left hand side of \eqn{eq:sigmaisolexp} is
independent of $\delta_0$, so must be the right hand side, too.  Below
we shall demonstrate that for sufficiently loose generation isolation,
in the range $\delta_0^{\gen}\in [0.01,0.1]$, the term
$\sigma_\incl\,\Theta_\isol^\gen\,\Theta_\isol^{\rm exp}$
obtained with usual experimental cone-type isolation of
\eqn{eqn:expisol}, depends on $\delta_0^{\gen}$ very little.
As a result, the second term on the right hand side of
\eqn{eq:sigmaisolexp} has to be almost independent of
$\delta_0^{\gen}$, too.  Although this second term is not computable in
perturbation theory, making $\Theta_\isol^\gen$ looser, it decreases,
and we expect it becomes negligible within the accuracy of the
calculation, when $\delta_0^{\gen} \leq 0.05$ and the photon is hard.
We shall discuss the accuracy of this assumption further in
\secref{sec:wgamma} by comparing the prediction from the first term
on the right hand side of \eqn{eq:sigmaisolexp} to the experimentally
measured cross section for a specific process.  This means that the
first term, $\sigma_\incl\,\Theta_\isol^\gen\,\Theta_\isol^{\rm exp}$
approximates the experimentally isolated hard photon cross section
within the accuracy of the prediction.


With such a generation isolation we can generate a sufficiently inclusive
sample of pre-showered events. On the events prepared this way it is easy to
apply a close-to-experiment type of cut such as \eqn{eqn:expisol}, the
quark-photon singularity is appropriately screened hence allowing for a
small hadronic (or partonic) activity in the cone around the photon and
cannot lead to infinite predictions. This procedure of making
theoretical predictions is made possible by the generation of LHEs as
opposed to producing differential distributions directly, as in the
case of computing cross sections at fixed order in perturbation theory
beyond LO accuracy.

\section{Independence of the generation isolation}
\label{sec:isolation-indep}

\begin{figure}
\includegraphics[width=0.50\textwidth]{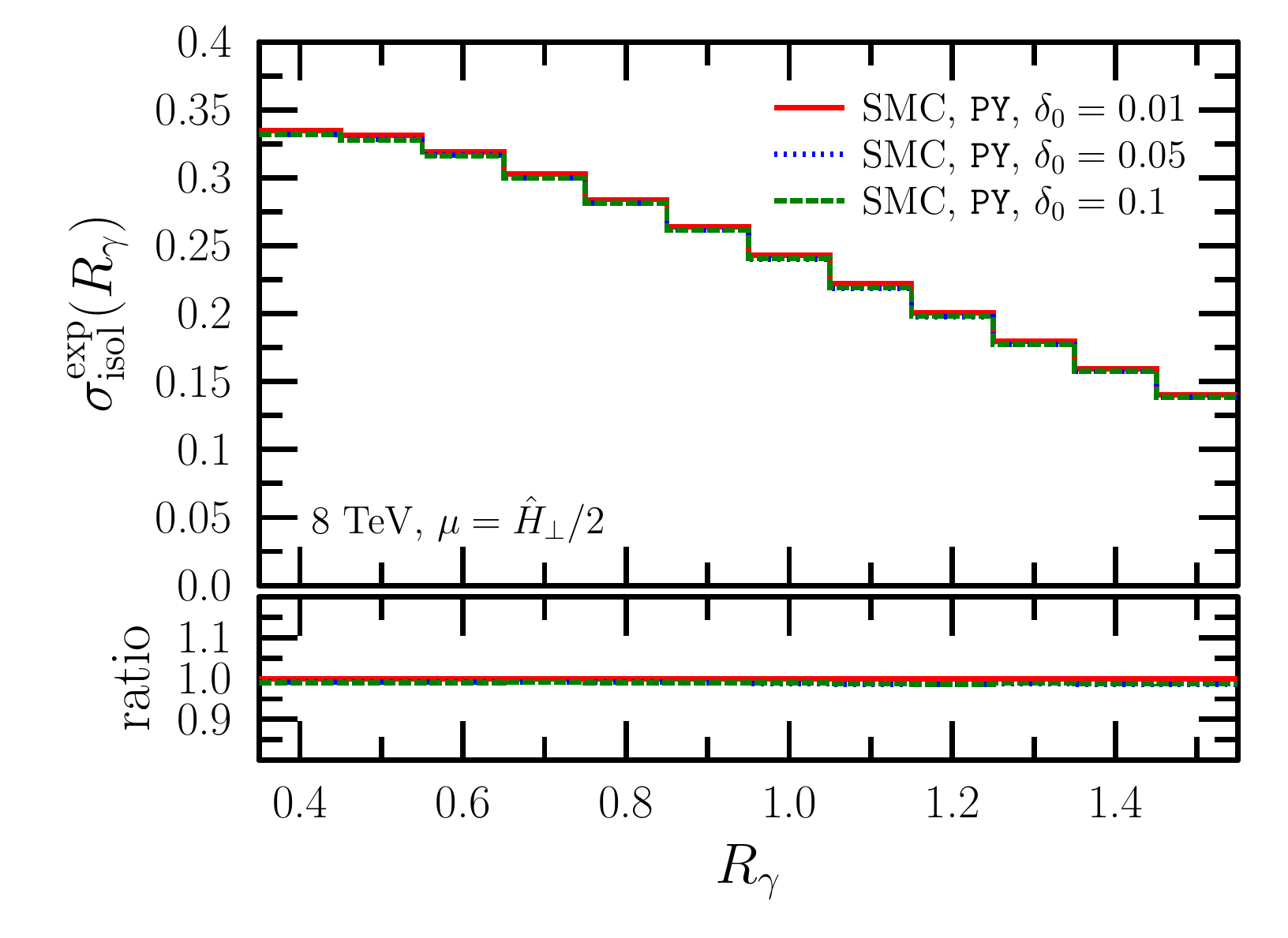}
\includegraphics[width=0.50\textwidth]{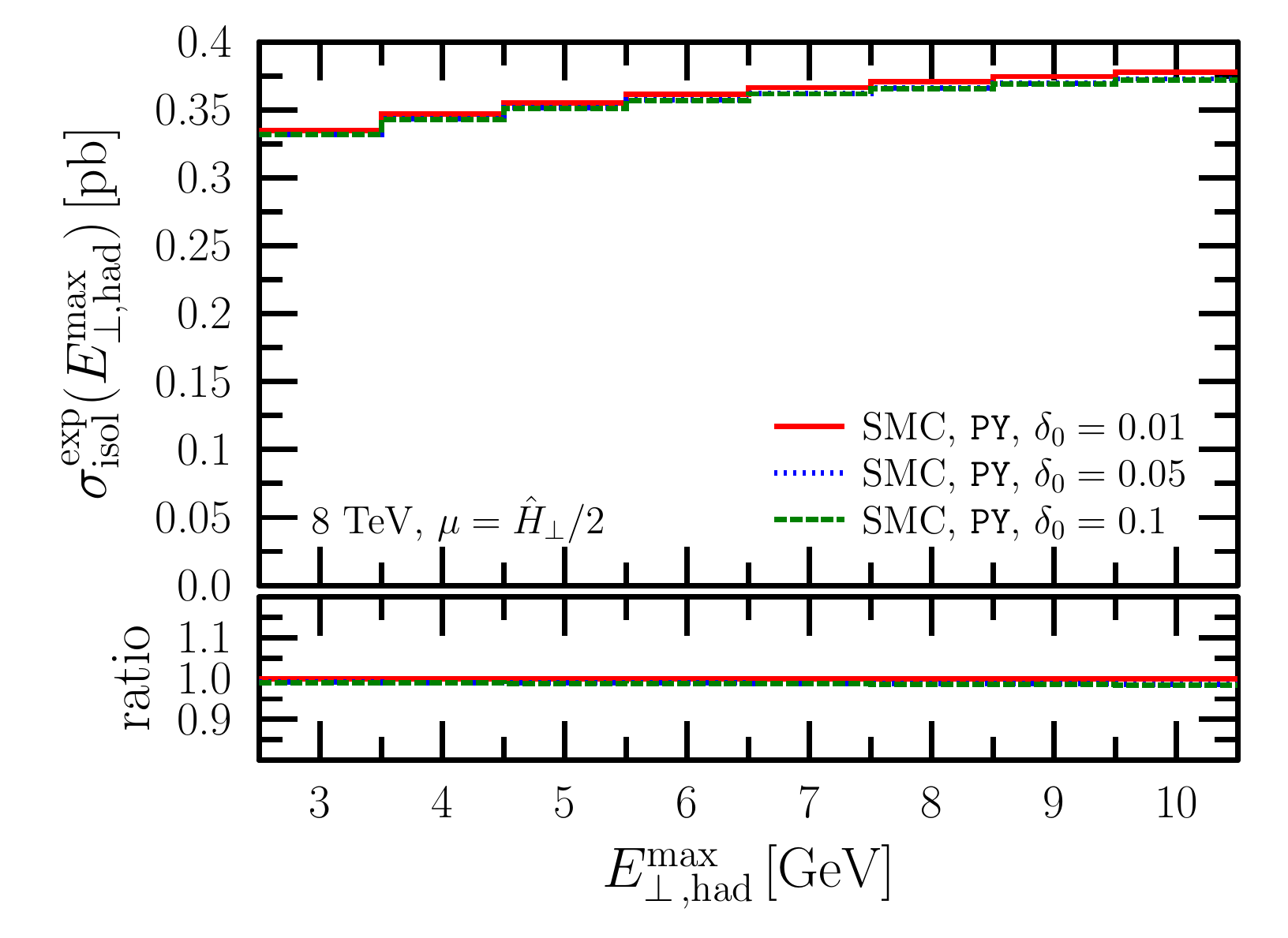}
\caption{\label{fig:SMC-sigmaR} 
Isoloted photon cross sections obtained after
full SMC with different generation isolations using cuts listed in the text,
as a function of
a) the radius of experimental photon isolation cone,
b) the hadronic leakage inside the photon isolation cone.}
\end{figure}

We generate an almost inclusive event sample with a loose photon
isolation. The generation isolation parameter $\delta_0^\gen$ should be
chosen such that the distributions with experimental photon isolation
obtained at various stages of event simulation (from LHEs, after parton
shower and after full SMC) should be independent of it. In order to see
this independence, we generated events with three different generation
isolation values: $\delta_0^\gen \in\{0.01,0.05,0.1\}$. These event
generations are done with parameters listed in \secref{sec:NLO-LHEcomp}. 
Then we compare the predictions made with different values of
$\delta_0^\gen$ at various stages of the event simulation.  Although
the particle content can be different at different stages of event
evolution, we kept the set of cuts applied to the events the same:
\begin{itemize}
\item{There is a cut on the transverse momentum of the hardest photon:
$\ptgamma > 30\,\gev$.}

\item{The hardest photon should be central: $|y_\gamma| < 2.5$.}

\item{A jet algorithm is applied using the anti-$\kt$ algorithm 
\cite{Cacciari:2008gp} provided by \fastjet\ 
\cite{Cacciari:2011ma,Cacciari:2005hq} with $\ptj > 30\,\gev$ and 
$R = 0.4$.}

\item{The hardest photon should be well-isolated from the jets:
$\Delta R(\gamma,j) > 0.4$ measured on the rapidity--azimuthal angle plane.}

\item{A hadronic (or partonic) leakage is allowed in an $R_\gamma=0.4$
cone around the photon according to \eqn{eqn:expisol} with 
$E_{\perp\,,{\rm had}}^{\rm max} = 3\,\gev$.}
\end{itemize}
We have checked that for $\delta_0^{\gen} \in [0.01,0.1]$ the
physical predictions depend marginally on $\delta_0^{\gen}$ at all
stages of the event evolution, but show here only for predictions
obtained at the hadronic stage, i.e.~after SMC. The cross section
values after full SMC and given selection cuts are presented as a
function of the radius of experimental photon isolation cone, and of
the hadronic leakage inside the photon isolation cone in
\fig{fig:SMC-sigmaR}. We see that independently of these parameters
(within the ranges shown here), the physical cross section depends
on the generation isolation weakly.

\begin{figure}
\includegraphics[width=0.50\textwidth]{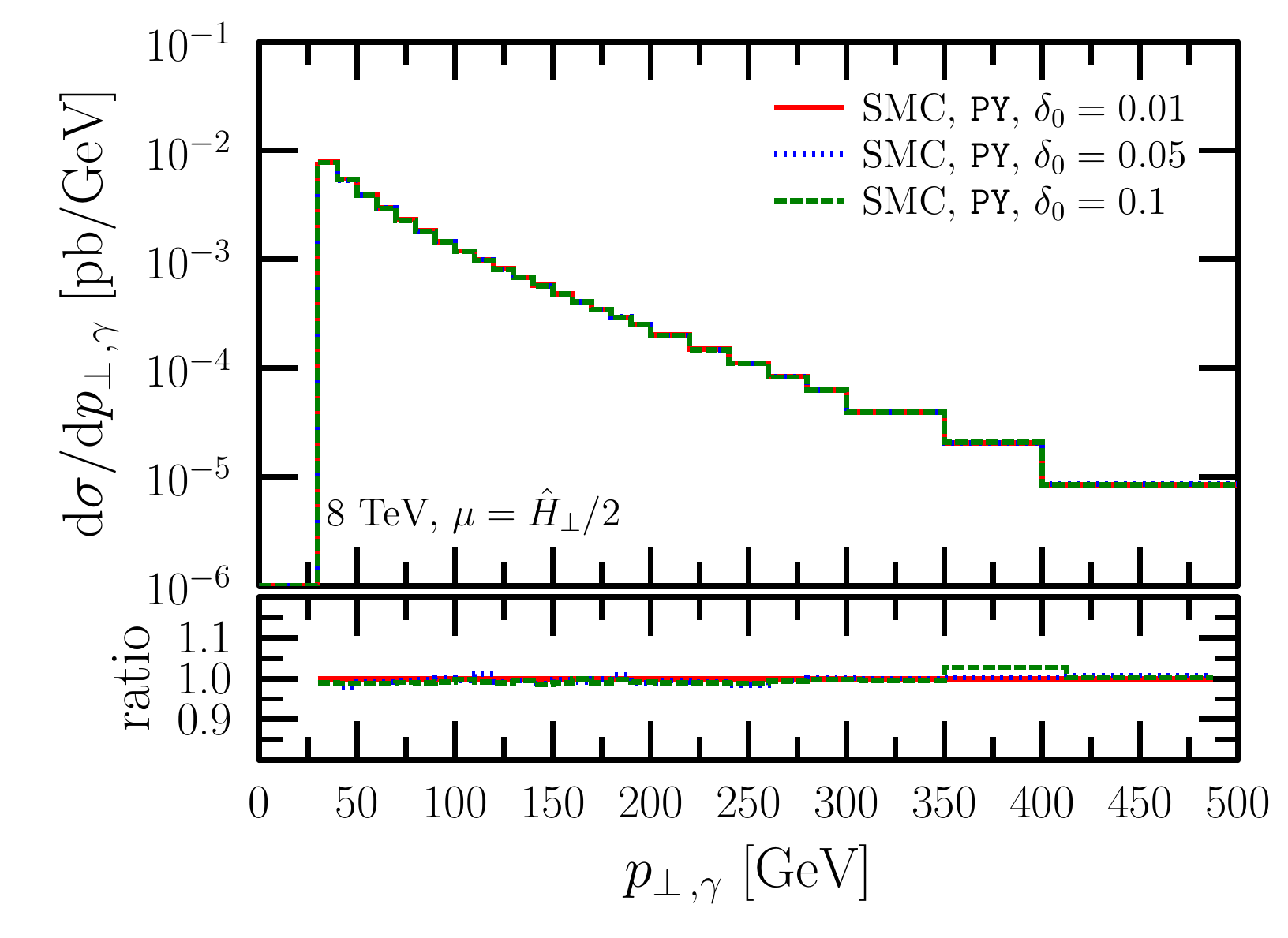}
\includegraphics[width=0.50\textwidth]{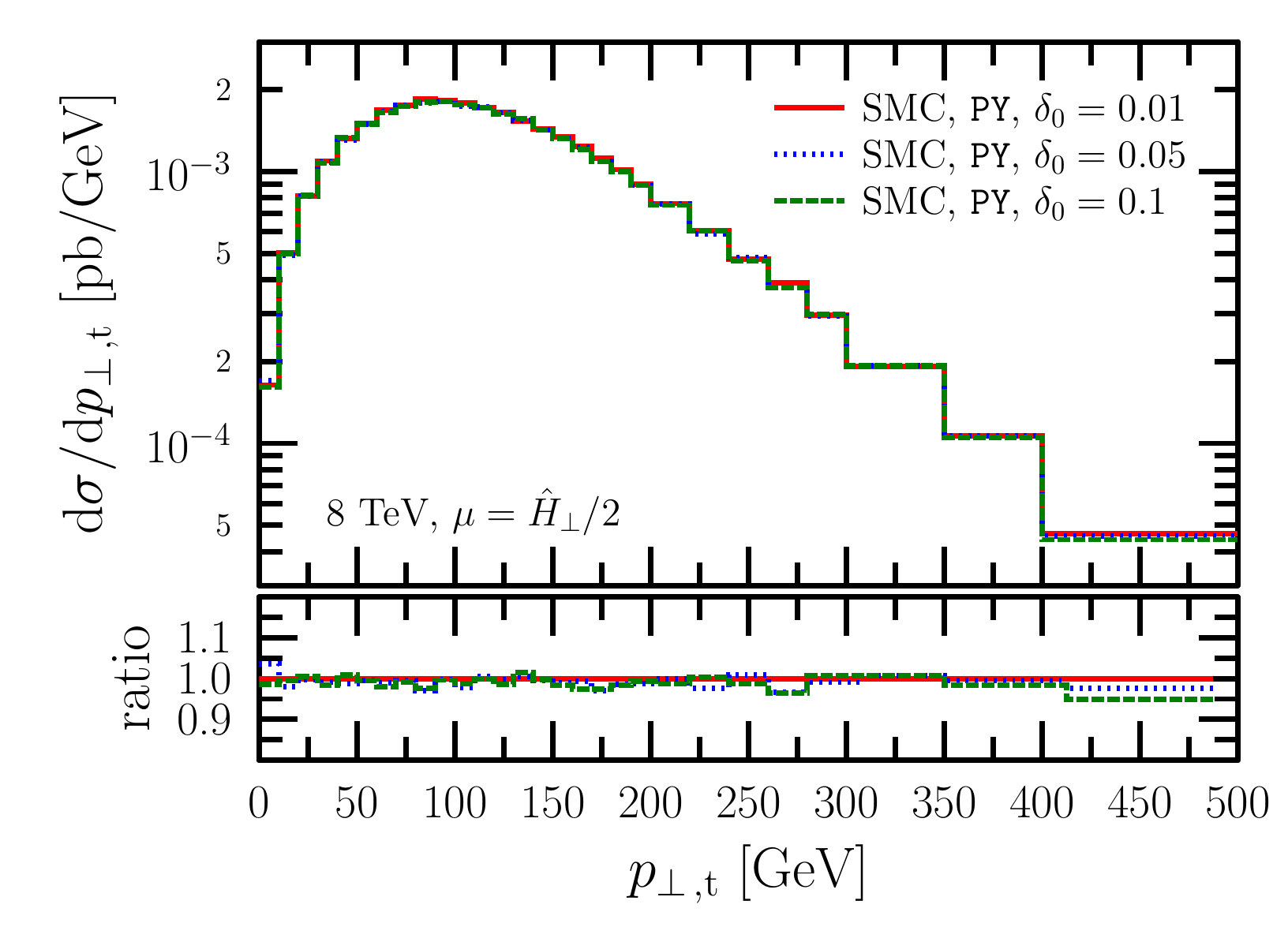}
\caption{\label{fig:SMC-pt} Transverse-momentum distribution for the
hardest photon and the t-quark after parton shower and hadronization with 
\pythia\ for smooth generation isolation with
$\delta_0^\gen \in [0.01,\,0.1]$.
On the lower panels the ratios of the predictions are shown.
}
\end{figure}

For kinematic distributions we find even smaller dependence on $\delta_0^\gen$.
Six sample distributions are presented on \figss{fig:SMC-pt}{fig:SMC-dR}.

\begin{figure}
\includegraphics[width=0.50\textwidth]{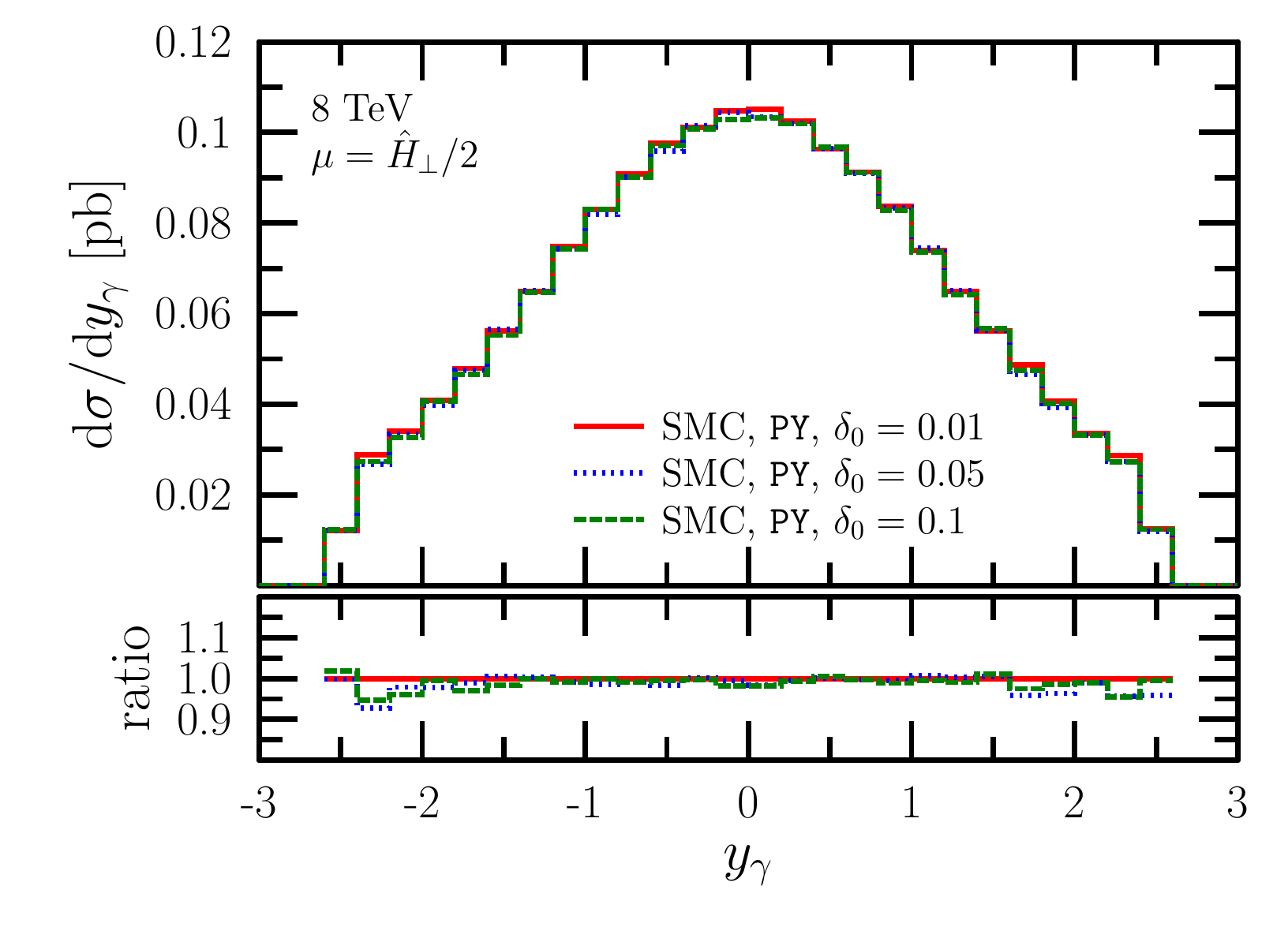}
\includegraphics[width=0.50\textwidth]{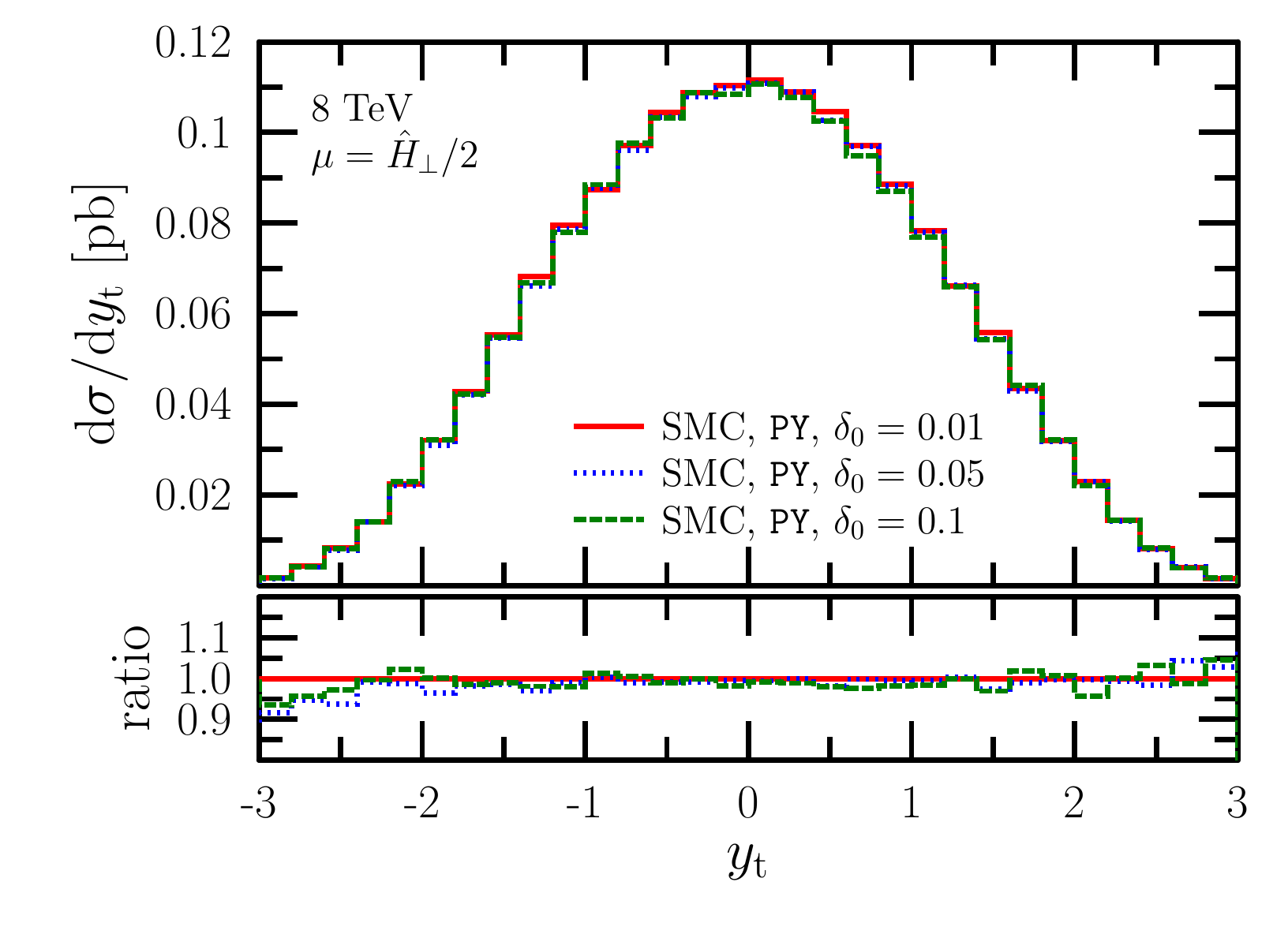}
\caption{\label{fig:SMC-y} The same as \fig{fig:SMC-pt} but for
the rapidities of the hardest-photon and the t-quark.}
\end{figure}

\begin{figure}
\includegraphics[width=0.50\textwidth]{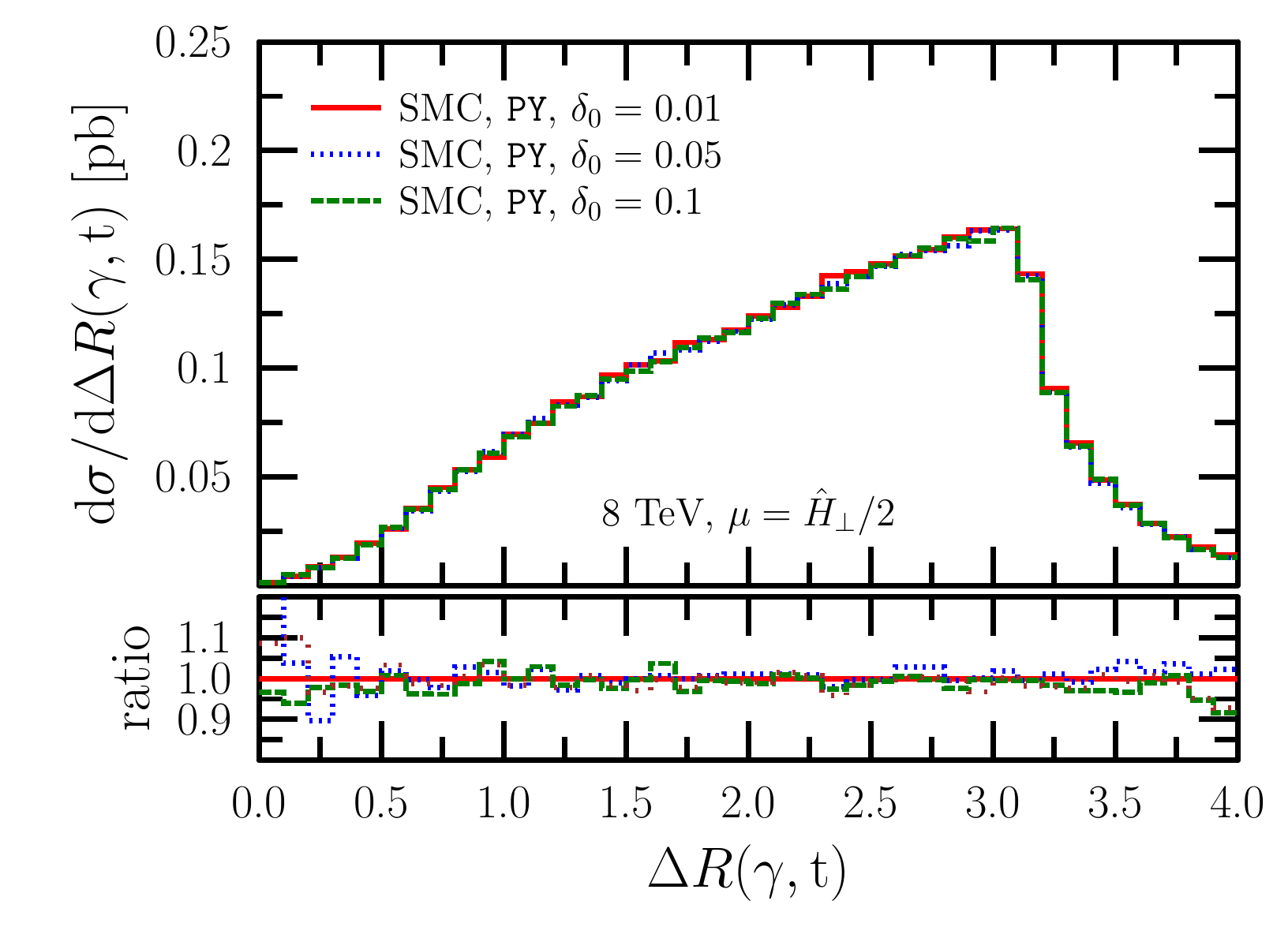}
\includegraphics[width=0.50\textwidth]{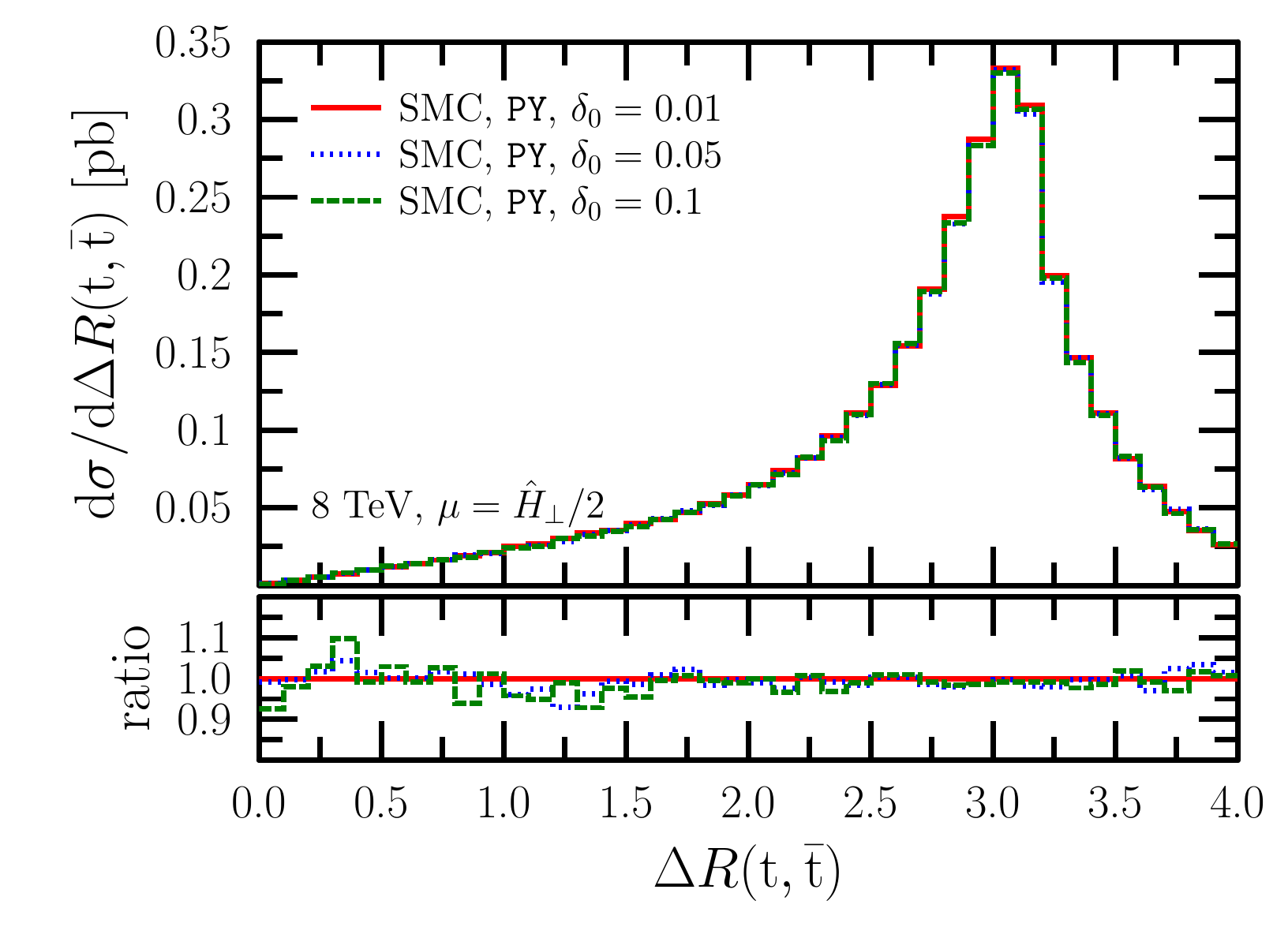}
\caption{\label{fig:SMC-dR} The same as \fig{fig:SMC-pt} but for
separations in the rapidity--azimuthal angle plane.}
\end{figure}

\section{Estimation of the non-pertrubative contribution}
\label{sec:wgamma}

In \eqn{eq:sigmaisolexp} we decomposed the isolated photon cross section
into a perturbatively computable part (first term) and a non-perturbative
contribution (second term). Furthermore, we argued that provided $\delta_0^\gen$
sufficiently small, we expect the non-perturbative contribution to be small.
This statement can only be verified by explicit comparison to experimental
data, which is presently not possible for isolated photon production in
association with a \ttbar~pair. It is possible however, for the case of
massive vector boson + isolated photon production for which the ATLAS
collaboration published results for both isolated photon + 0 jet
(exclusive) and isolated photon +$N (\geq 0)$ jets (inclusive) in the final
state \cite{Aad:2013izg}. This final state has also been considered
recently at NLO accuracy interfaced to a shower generator according to
the POWHEG prescription supplemented with the MiNLO procedure
\cite{Barze:2014zba}.  In this work the fixed order result is matched
to an interleaved QCD+QED parton shower, in such a way that the
contribution arising from hadron fragmentation into photons is fully
modeled. Thus for this process the comparison is possible not only for
experimental results, but also with a theoretical prediction where the
fragmentation is included through a shower model.

Within \powhel\ the $W\gamma$ process can be implemented
straightforwardly. We generated events for $W\gamma$ production with
the same three values of $\delta_0^\gen$ as in the case of \ttgamma\
production and checked that the predictions from the pre-showered
events agree with those at NLO accuracy, just as in the case of \ttgamma\
production in
\secref{sec:NLO-LHEcomp}. Next, we checked the dependence on the
generation isolation parameter, similarly as in
\secref{sec:isolation-indep} and found that the perturbative prediction
depends weakly (below 10\,\% for the exclusive and below 5\,\% for the
inclusive case) on the choice of $\delta_0^\gen$ in
the range $\delta_0^\gen \in [0.01,0.1]$ if a similar physical isolation
is used as in \secref{sec:isolation-indep}. Thus we have implemented
the event selection of ref.~\cite{Aad:2013izg} and made predictions for the
inclusive case and for the exclusive case using events obtained with
$\delta_0^\gen = 0.05$. 

\begin{figure}
\includegraphics[width=0.50\textwidth]{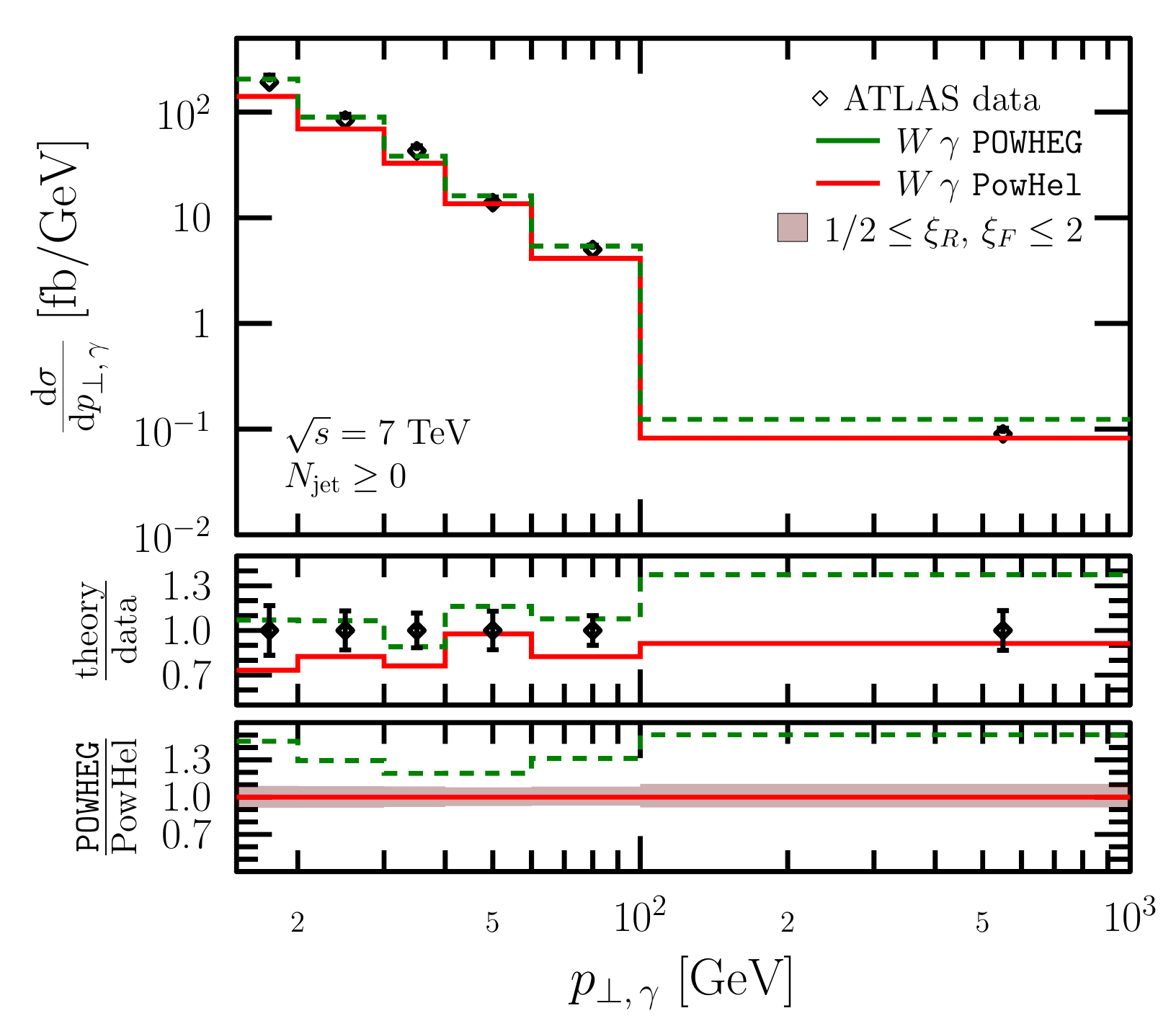}
\includegraphics[width=0.50\textwidth]{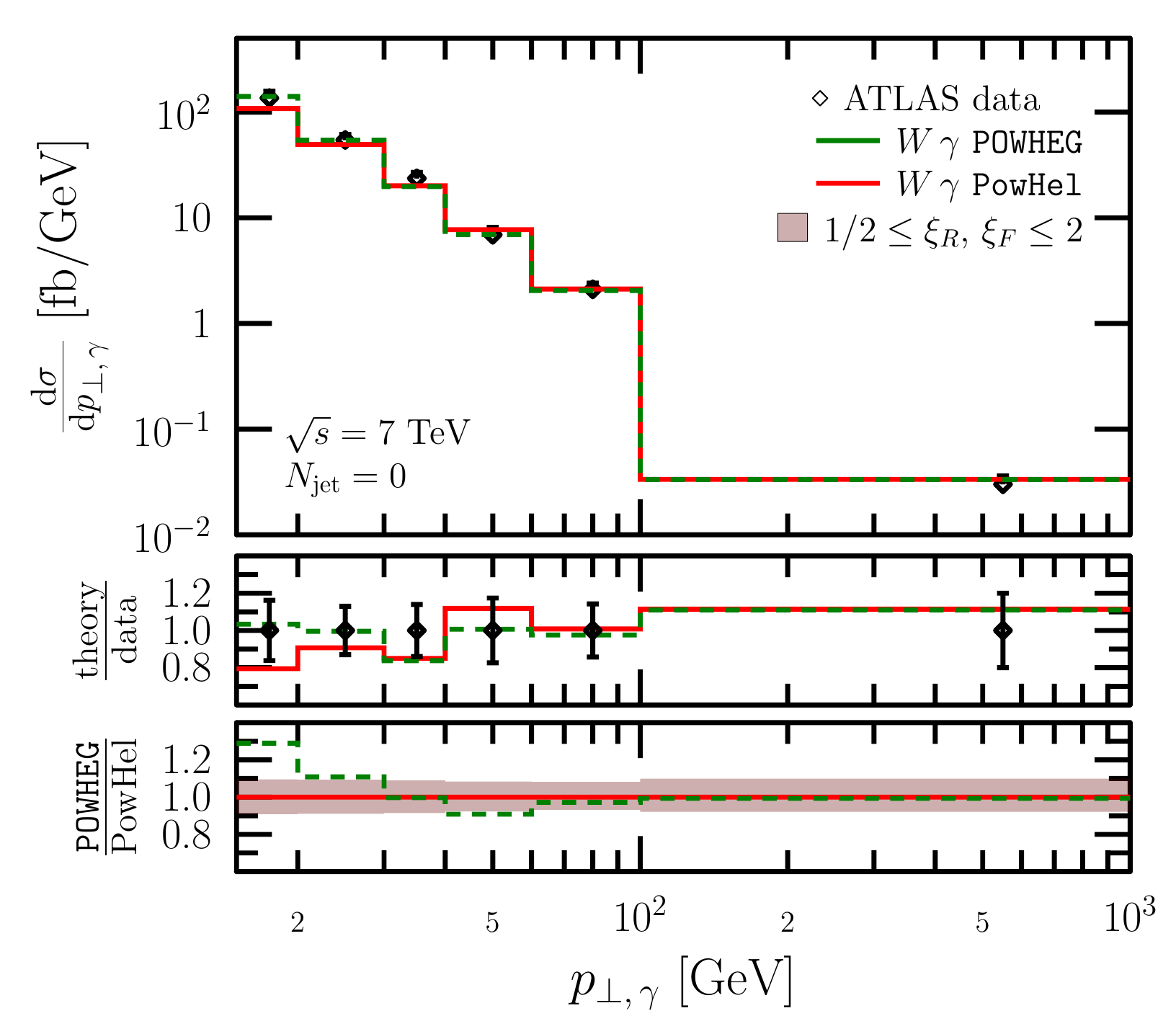}
\caption{\label{fig:pTgamma-Wgamma} Transverse momentum distribution of
the isolated photon for the
a) $W + \gamma + N \geq 0$ jets and b) $W + \gamma + 0$ jet final states.
The lower panels show the ratio of the predictions to the data.}
\end{figure}

We show our predictions for
the transverse momentum distribution of the isolated photon, compared
to the predictions of ref.~\cite{Barze:2014zba} and the results of
ref.~\cite{Aad:2013izg} in \fig{fig:pTgamma-Wgamma}.

In general, we find that our approach gives a good description of the
data if the radiated photon is harder than the accompanying jets. Thus
for the exclusive case, the data overlap with our predictions for the
full range within the uncertainty of the prediction. In fact, for this
case the two theoretical predictions also coincide within the scale
dependence band except for the first bin.  For the inclusive case the
two predictions differ and the difference, in principle, may be
attributed to the neglected fragmentation contribution.  However, it
should be noted that the predictions of ref.~\cite{Barze:2014zba} do
not use non-perturbative information extracted from data, but a model of
fragmentation.  At the present accuracy of the data it is difficult to
make a clear conclusion which prediction is favoured by experiment, but
the following general trend seems to emerge: the harder the photon the
better the agreement between our prediction and the data, while for the
case of matched NLO to an interleaved QCD+QED parton shower, the
agreement is better for small transverse momenta, with transition
around 60\,GeV for the given selection cuts.

\section{Effect of the parton shower}

\begin{figure}
\includegraphics[width=0.50\textwidth]{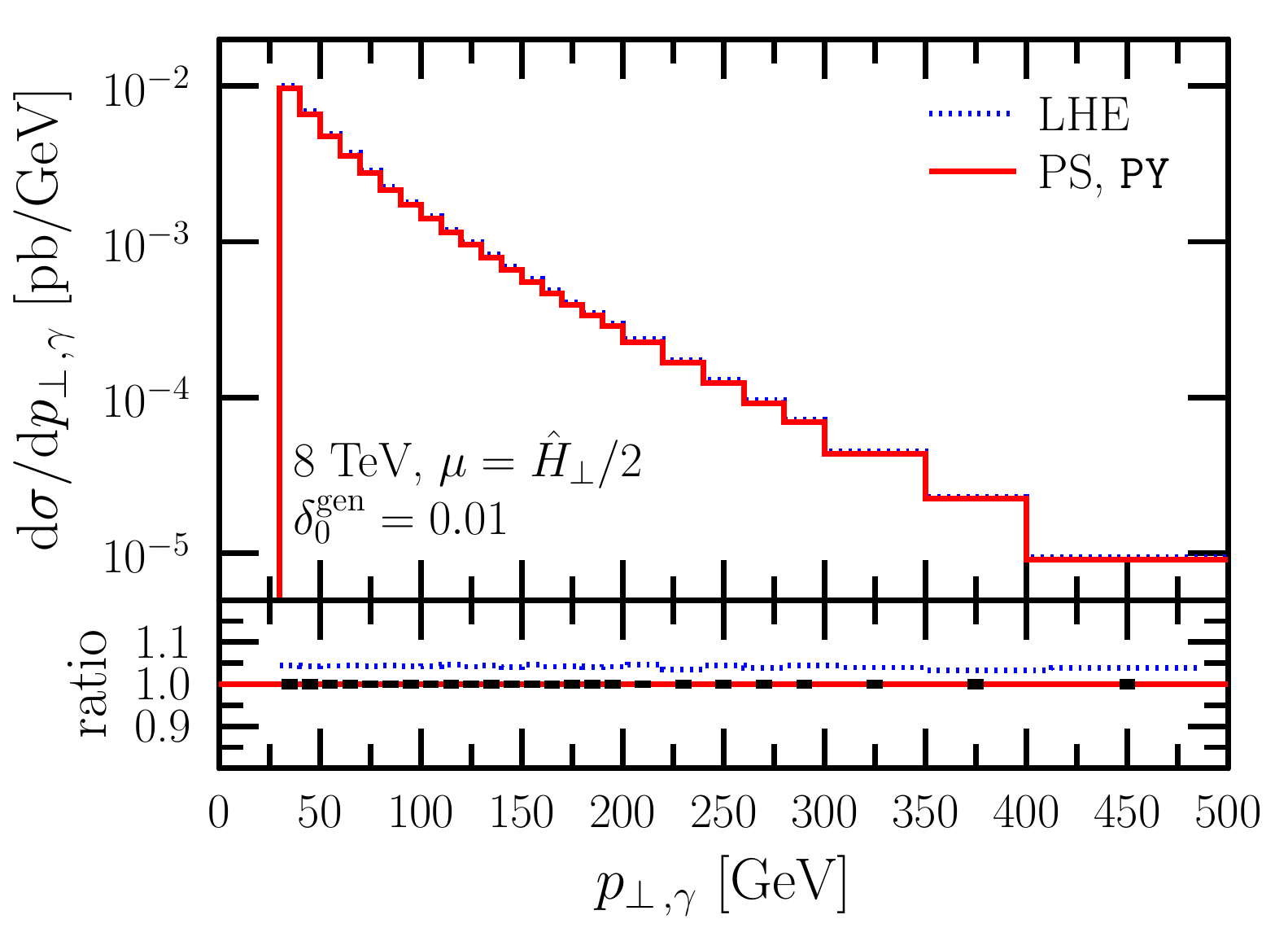}
\includegraphics[width=0.50\textwidth]{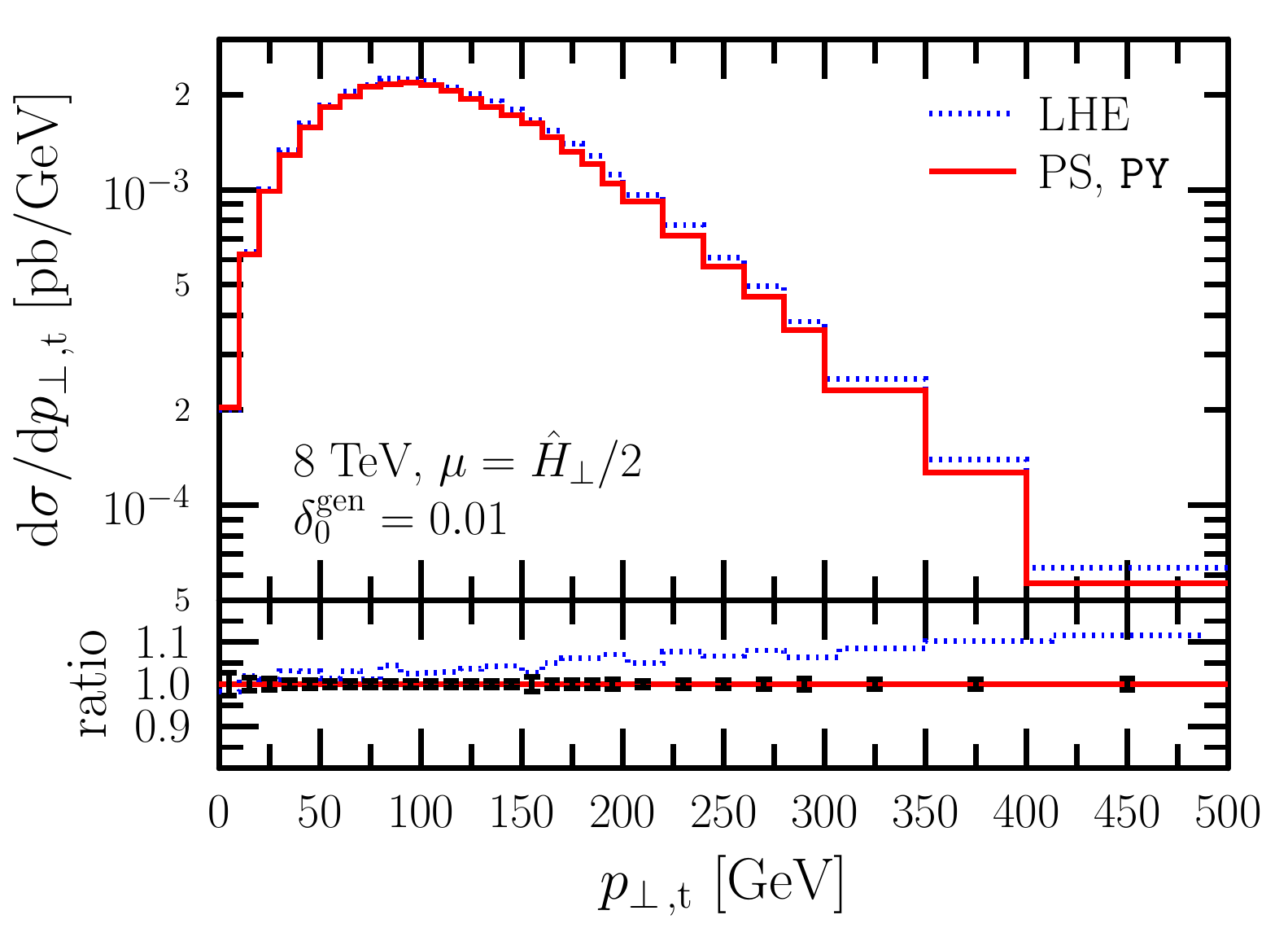}
\caption{\label{fig:LHE-PS-pt} Transverse-momentum distribution for the
photon and t-quark at the LHE stage and after parton shower. The lower
panel shows the LHE/PS ratio.}
\end{figure}

In the previous sections we estimated the effect of the neglected
non-perturbative contribution, as well as demonstrated that predictions
for isolated photon cross section made with full SMC do not depend upon
the sufficiently small generation-isolation. In fact, we also checked 
that the latter is true at various stages of event simulation (LHE, PS
and SMC). To quantify the effect
of the parton shower and in the next section to present physical
predictions after full SMC we decided to use $\delta_0^{\gen} = 0.01$
in our generation isolation. For this comparison we used the setup of
\secref{sec:isolation-indep}. Our standard distributions can be found on
\figss{fig:LHE-PS-pt}{fig:LHE-PS-dR}. While for rapidities and
separations the difference between the LHE and PS stages only manifest
in an overall change in normalization, for the transverse-momentum
distributions the change is not only a constant factor in
normalization, but there is even a change in the shape. As we expect, the
shower softens the spectra. This softening added to the difference
between the predictions of LHEs and at NLO suggests very small PS
effect at high transverse momenta. (We cannot compare
\figs{fig:LHE-NLOcompFx-pta-pttq}{fig:LHE-PS-pt} directly as photon
isolations are different.) In the case of the photon \pt\ the change
remains small, around 5\%, while
for the transverse momentum of the t-quark it reaches even 12\% when
the \pt\ approaches $500\,\gev$. If our default, rather tight,
criterion on the allowed hadronic leakage is loosen up (going from
$3\,\gev$ to $10\,\gev$) the difference observed in the photon
transverse-momentum distribution remains more-or-less the same, but in
the case of the transverse momentum of the t-quark the difference drops
below 10\% in the high-\pt\ region.  The relaxation in
the hadronic leakage condition results in a smaller difference,
$\sim 1\%$, for rapidities and separations.

\begin{figure}
\includegraphics[width=0.50\textwidth]{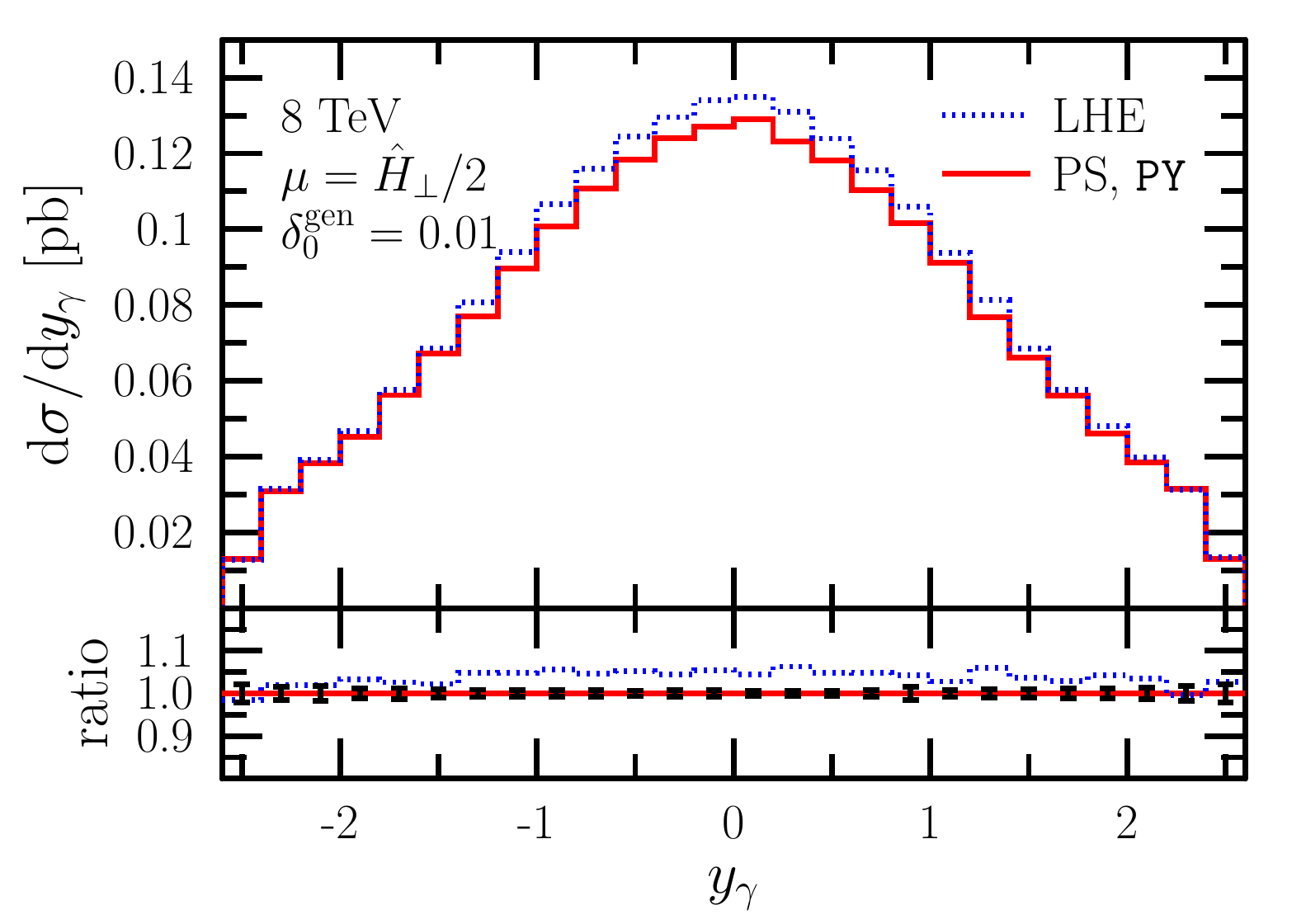}
\includegraphics[width=0.50\textwidth]{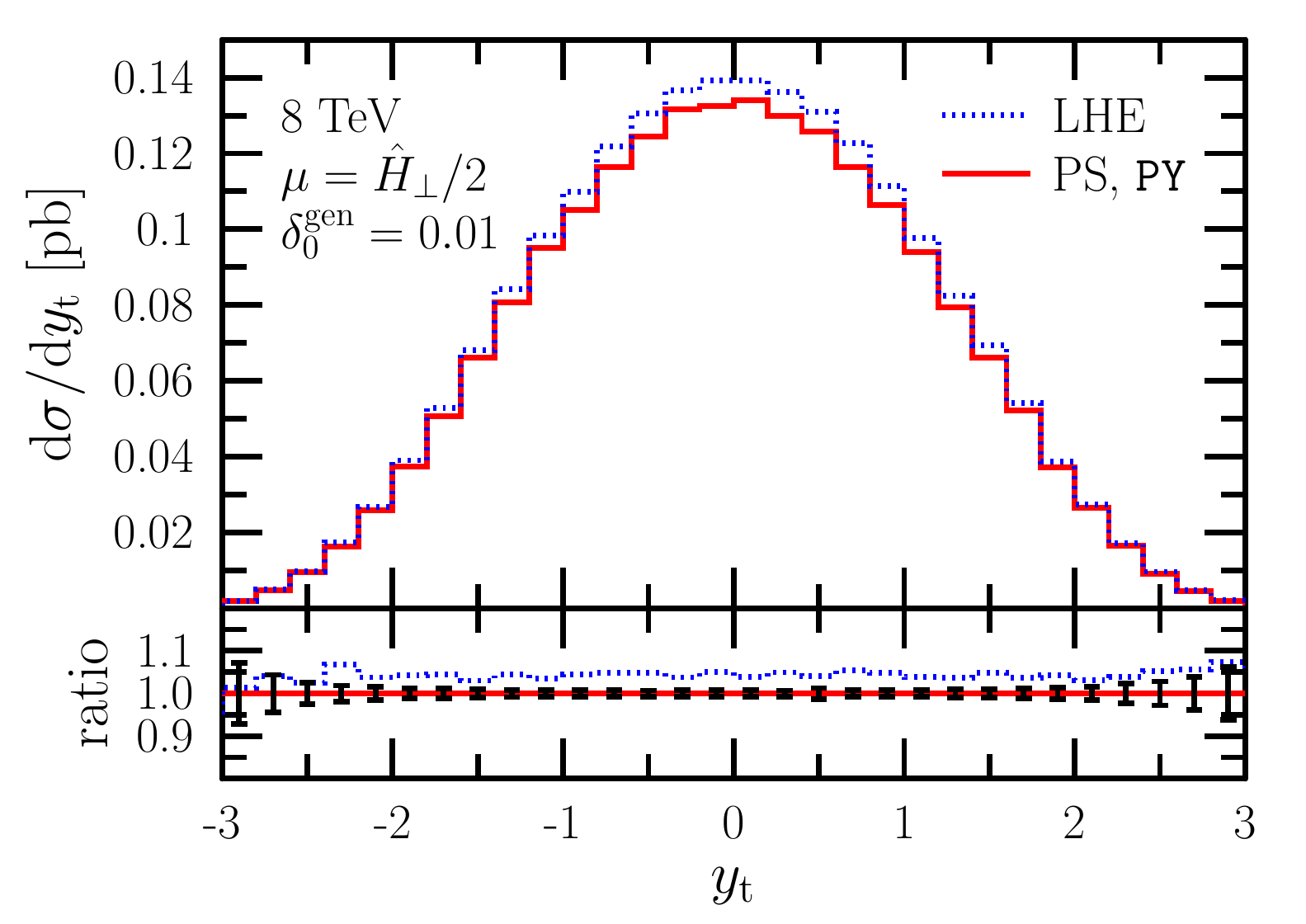}
\caption{\label{fig:LHE-PS-y} The same as \fig{fig:LHE-PS-pt} but for the 
rapidities of the photon and t-quark.}
\end{figure}

\begin{figure}
\includegraphics[width=0.50\textwidth]{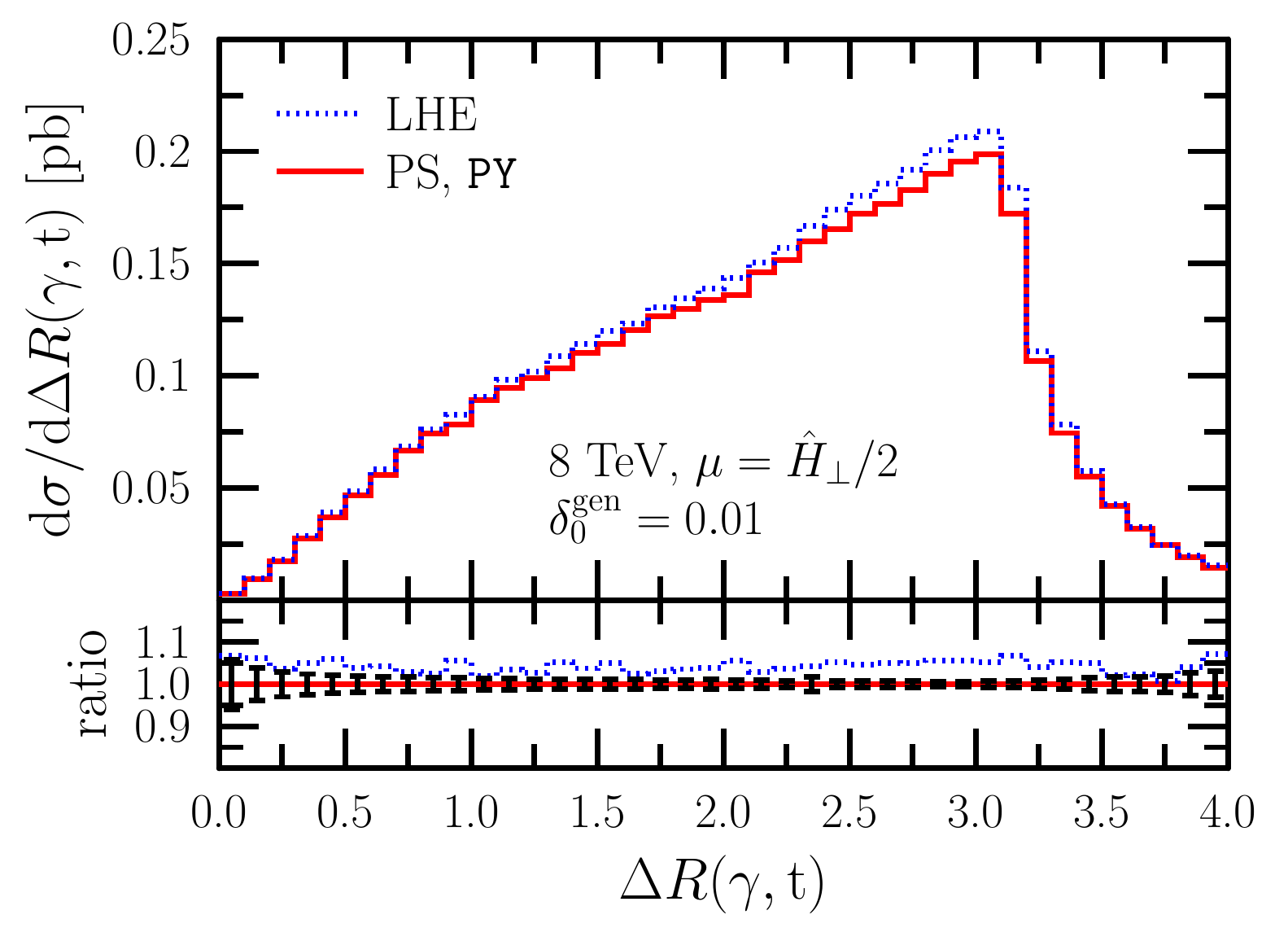}
\includegraphics[width=0.50\textwidth]{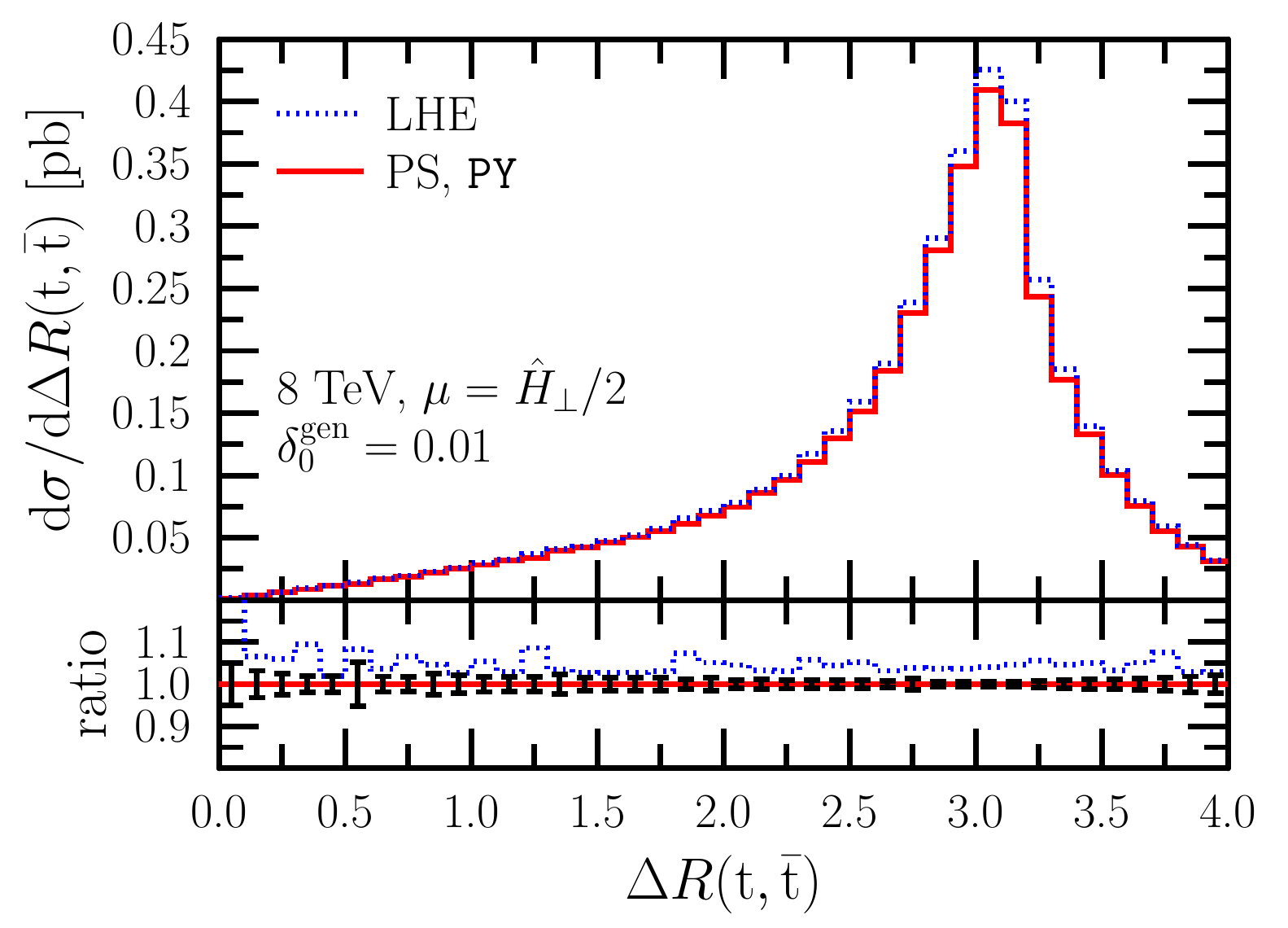}
\caption{\label{fig:LHE-PS-dR} The same as \fig{fig:LHE-PS-pt} but for the 
separation of the photon-t and \ttbar\ systems.}
\end{figure}

\section{Predictions}

We conclude with a simple phenomenological study at the 
hadron level. To this end \pythiaver\ was chosen to decay, shower and
finally hadronize the events. The event sample with $\delta_0^\gen = 0.01$
at 8~TeV was selected, \pythia\ was run with the 2010 Perugia tune
\cite{Skands:2010ak}, omitting photon showers, making $\tau^\pm$ and
$\pi^0$ stable and we turned off multi-particle interactions.
The cuts employed in this analysis were the following:
\begin{itemize}
\item The analysis was done in the semileptonic decay-channel by
requesting exactly one hard lepton or antilepton in the final state
with $p_{\bot\,,\ell} > 30\,\gev$, the (anti)lepton had to be isolated from all
the jets with $\Delta R(\ell,j) > 0.4$.

\item The final state had to contain one hard photon in the central
region, $|y_{\gamma}| < 2.5$ with $\ptgamma > 30\,\gev$, isolated from all
the jets by $\Delta R(\gamma,j) > 0.4$. A minimal hadronic leakage was 
allowed in a $R_\gamma = 0.4$ cone around the photon with
$E_{\perp\,,{\rm had}}^{\rm max} = 3\,\gev$ according to \eqn{eqn:expisol}.

\item The (anti)lepton and photon had to be separated from each other, 
$\Delta R(\gamma,\ell) > 0.4$.

\item Jets were reconstructed with the anti-\kt\ algorithm 
\cite{Cacciari:2008gp} with $R = 0.4$ and $p_{\bot}^{j} > 30\,\gev$.

\item The event had to have significant missing transverse momentum,
$\ptmiss > 30\,\gev$.
\end{itemize}

In our calculation, throughout, a different scale choice was used
than that in the literature \cite{Melnikov:2011ta} for \ttgamma\
production.  Our default scale choice, the half the sum of transverse
masses $\hat{H}_\perp/2$ was already motivated in
\cite{kardos:2013vxa}. To see the difference between the two scale
choices a scale-uncertainty study is performed and scale-uncertainty
bands are shown for the distributions obtained at the hadron level. The
renormalization and factorization scales are defined as $\mur = \xi_R
\mu_0$ and $\muf = \xi_F \mu_0$, respectively, and the band is formed
as the upper- and lower-bounding envelopes of distributions taken with
\begin{align}
(\xi_R,\xi_F) 
\in
\left\{
\left(\frac{1}{2},\frac{1}{2}\right),
\left(\frac{1}{2},1\right),
\left(1,\frac{1}{2}\right),
\left(1,1\right),
\left(1,2\right),
\left(2,1\right),
\left(2,2\right)
\right\}
\,.
\end{align}
The antipodal choices ($(1/2,2)$ and $(2,1/2)$) are left out. When
these are included, the uncertainty band for rapidities and separations
are unchanged while for transverse momenta in the large
transverse-momentum region the band widens by a few percent.

\begin{figure}
\includegraphics[width=0.50\textwidth]{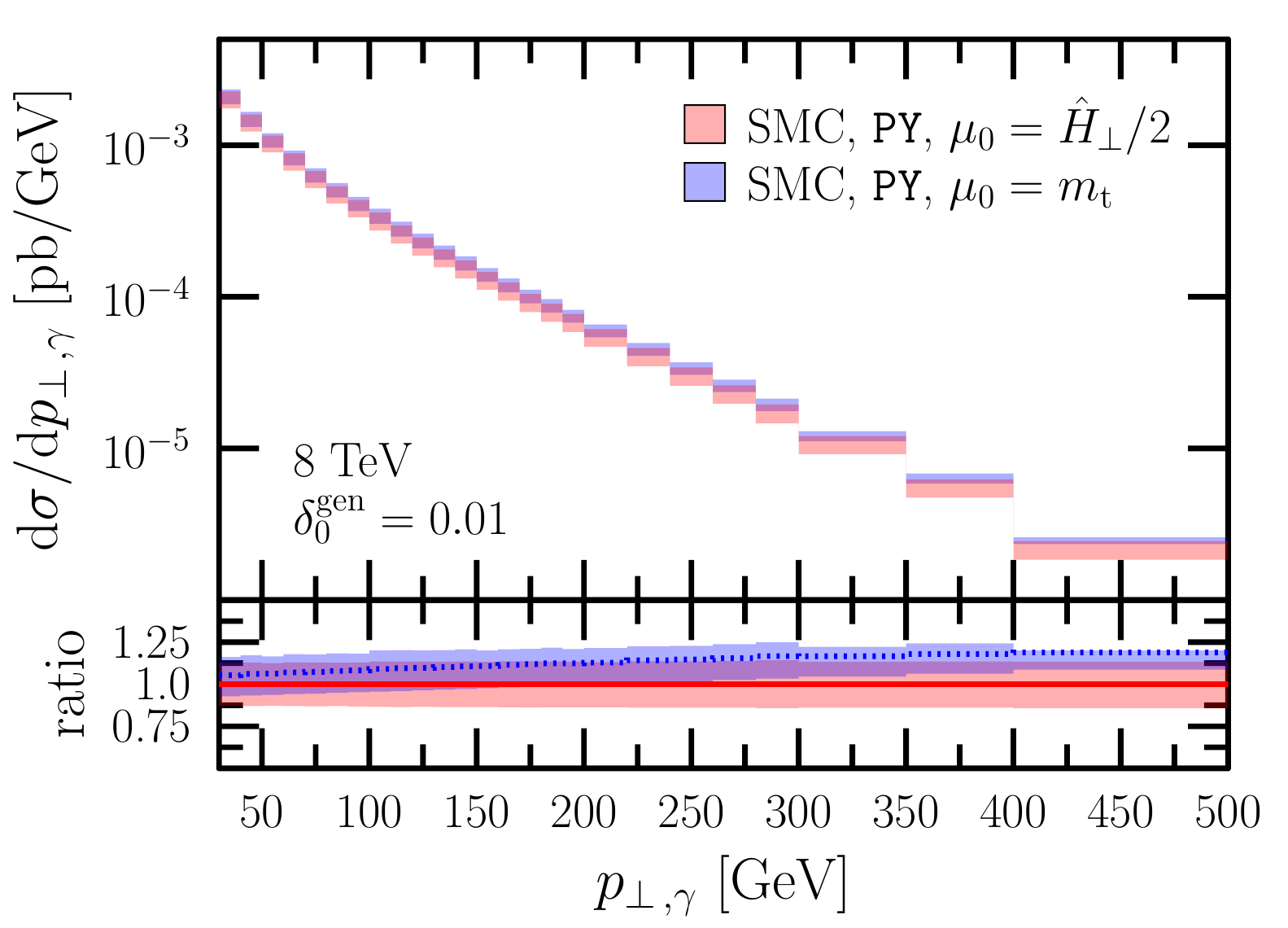}
\includegraphics[width=0.50\textwidth]{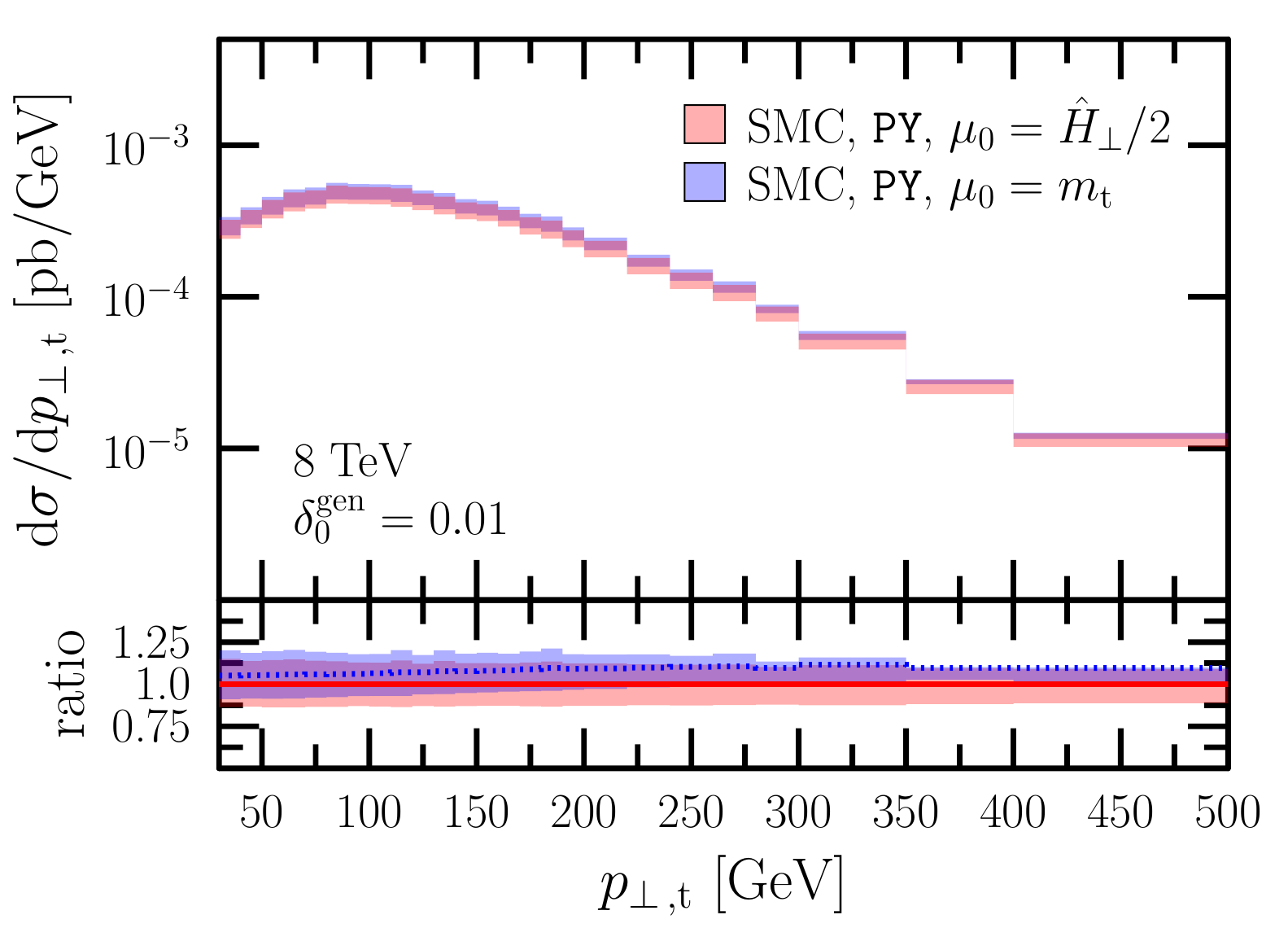}
\caption{\label{fig:SMCpred-pt-gamma-tq} Transverse momentum distribution 
for the photon and the t-quark at the hadronic stage. On the lower
panel the ratio of predictions to that obtained with our default scale
choice is shown.}
\end{figure}

In \fig{fig:SMCpred-pt-gamma-tq} the transverse momenta of the photon
and the t-quark are shown. The momentum of the t-quark is reconstructed
just like in the previous cases using \texttt{MCTRUTH}. Taking a look
at the transverse momentum of the photon the static scale results in a
narrower band with a shrinking width. This hints a cross-over point
at a higher \pt\ value, while in the case of the dynamical scale the
band, although wider, keeps the same width all across the whole plotted
transverse momentum spectrum. While for the \pt-distribution of the
photon the presence of a cross-over point is only hinted by the narrowing
uncertainty band, for the transverse momentum of the t-quark it is
indeed visible around 350\,GeV. Until this point the uncertainty band
taken with the static scale decreases in width than after opens up.
This is somehow expected since a highly boosted t-quark with a heavy
companion anti-t and a photon correspond to a system with a large
summed transverse mass hence lying far away from the central scale $m_\tq$.
\begin{figure}
\includegraphics[width=0.50\textwidth]{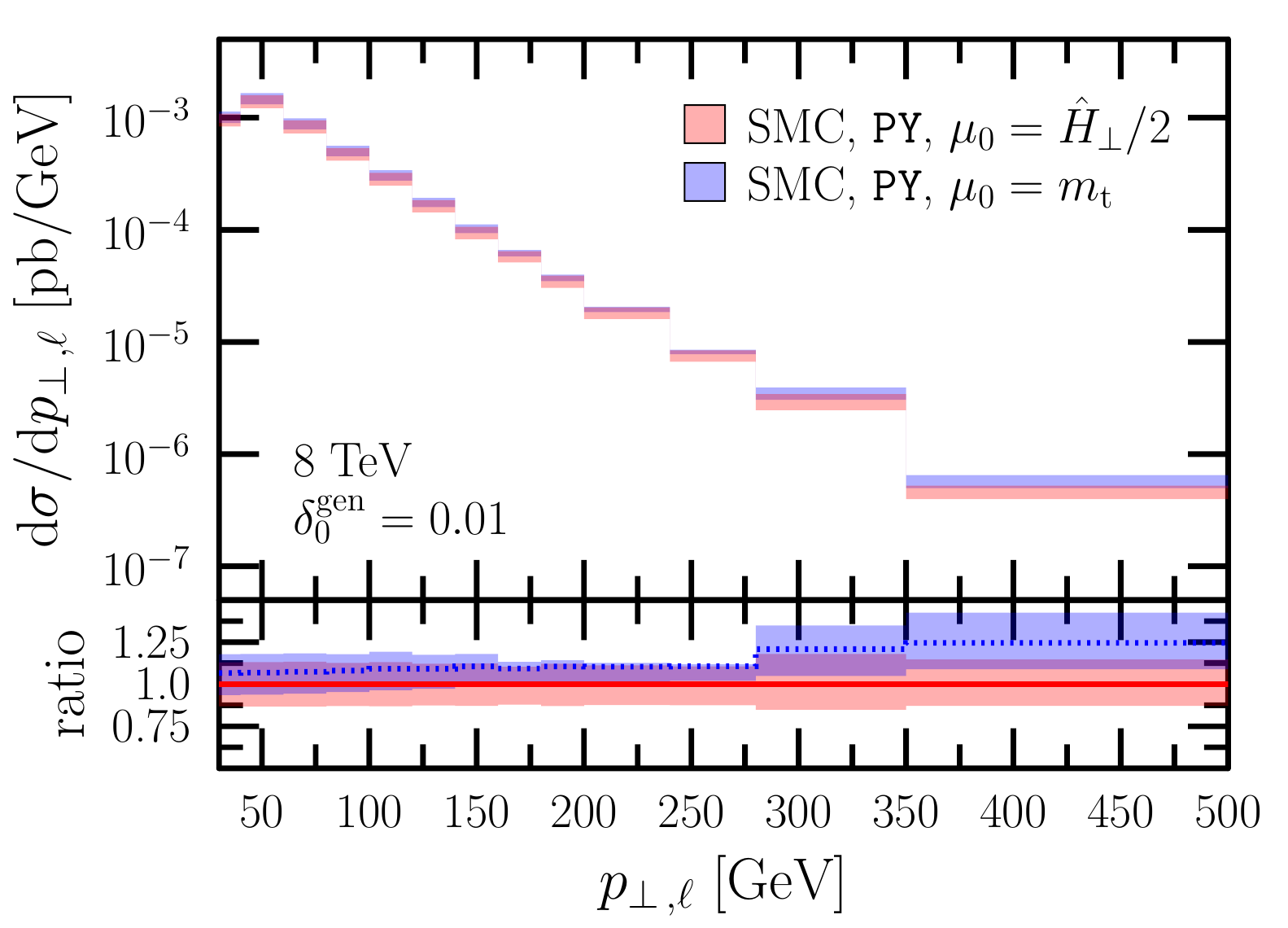}
\includegraphics[width=0.50\textwidth]{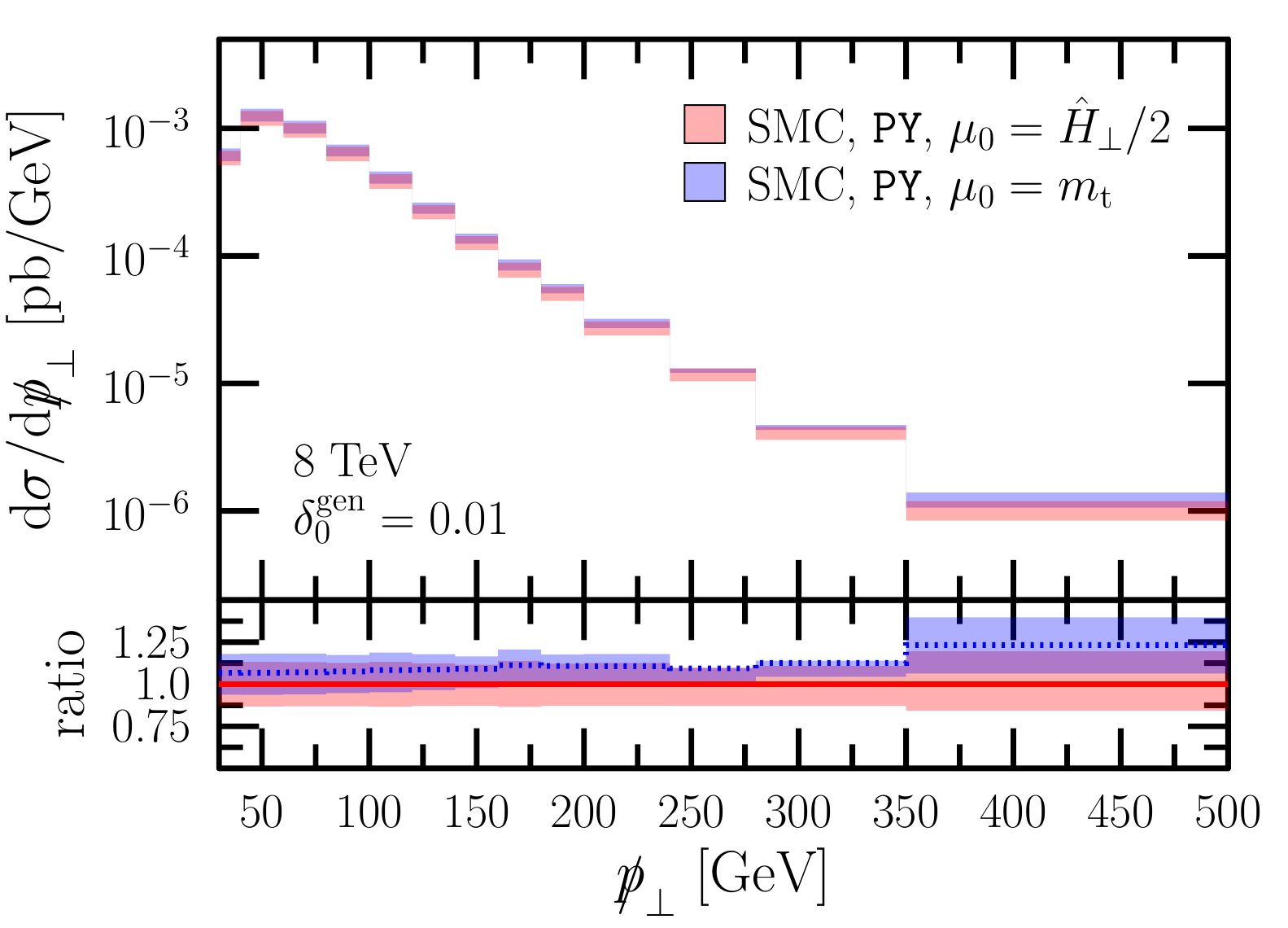}
\caption{\label{fig:SMCpred-pt-lept-miss} The same as 
\fig{fig:SMCpred-pt-gamma-tq} but for the spectra of the transverse
momentum of the charged (anti)lepton and the missing momentum.}
\end{figure}

In \fig{fig:SMCpred-pt-lept-miss} the spectrum of the transverse
momentum of the charged lepton and that of the missing momentum are
shown. For both distributions a cross-over can be seen around 250\,GeV
when static scale is used. The dynamical scale choice appears to give
reliable scale dependence over the whole plotted range for these
observables.

\begin{figure}
\includegraphics[width=0.50\textwidth]{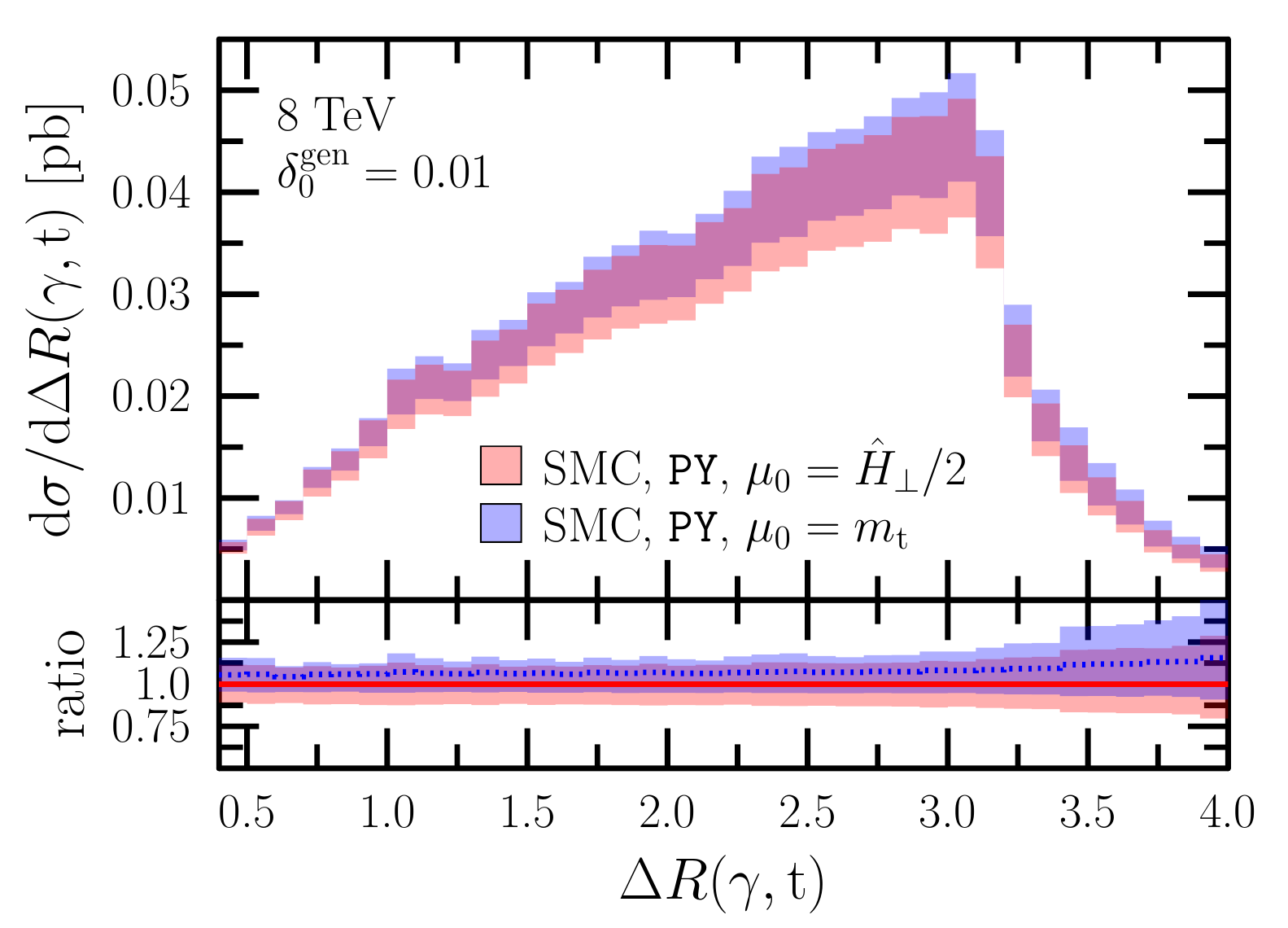}
\includegraphics[width=0.50\textwidth]{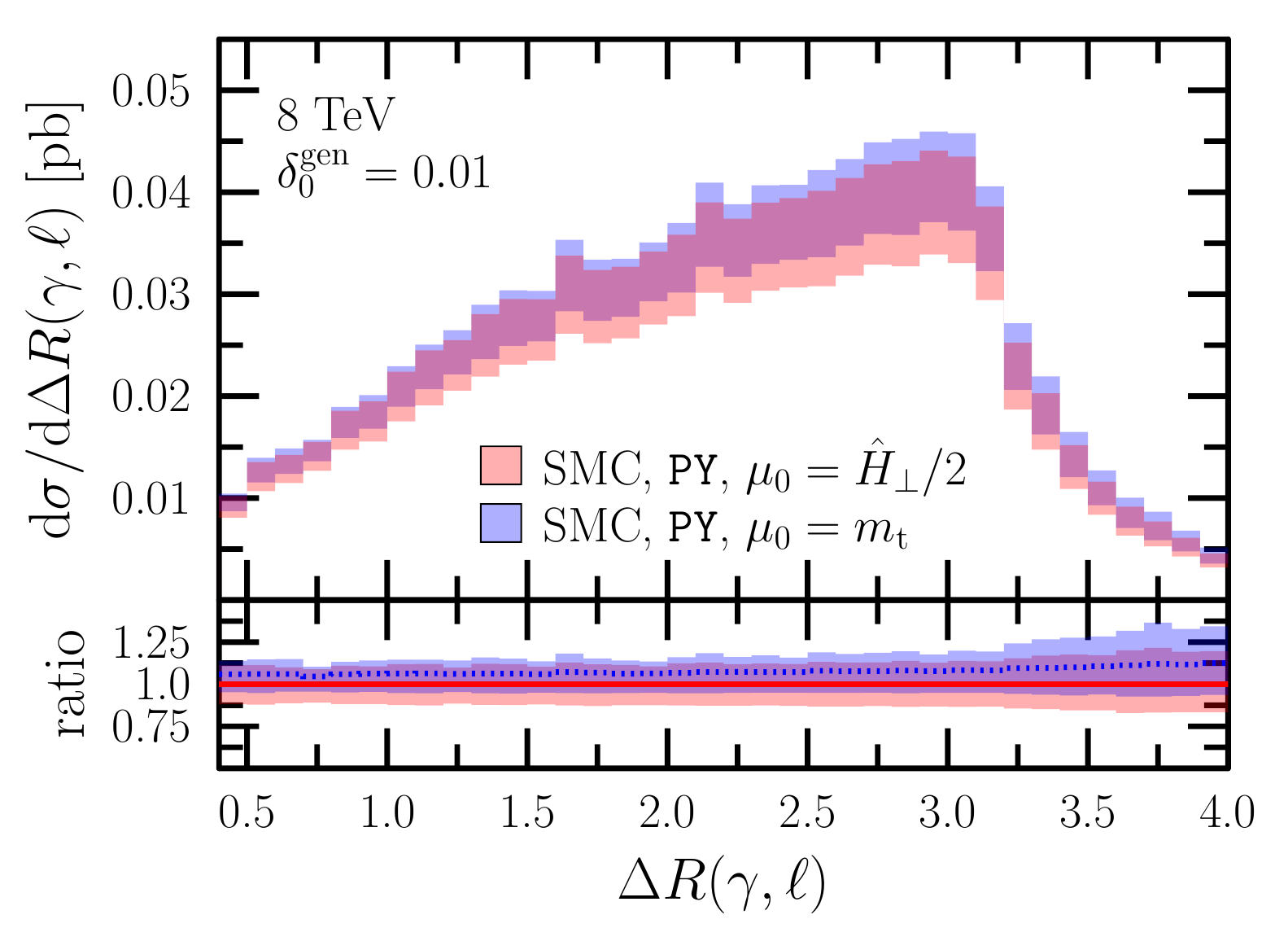}
\caption{\label{fig:SMCpred-dR-gamma-tq-lept} The same as 
\fig{fig:SMCpred-pt-gamma-tq} but for separations measured in the
rapidity--azimuthal angle plane.}
\end{figure}

\begin{figure}
\includegraphics[width=0.50\textwidth]{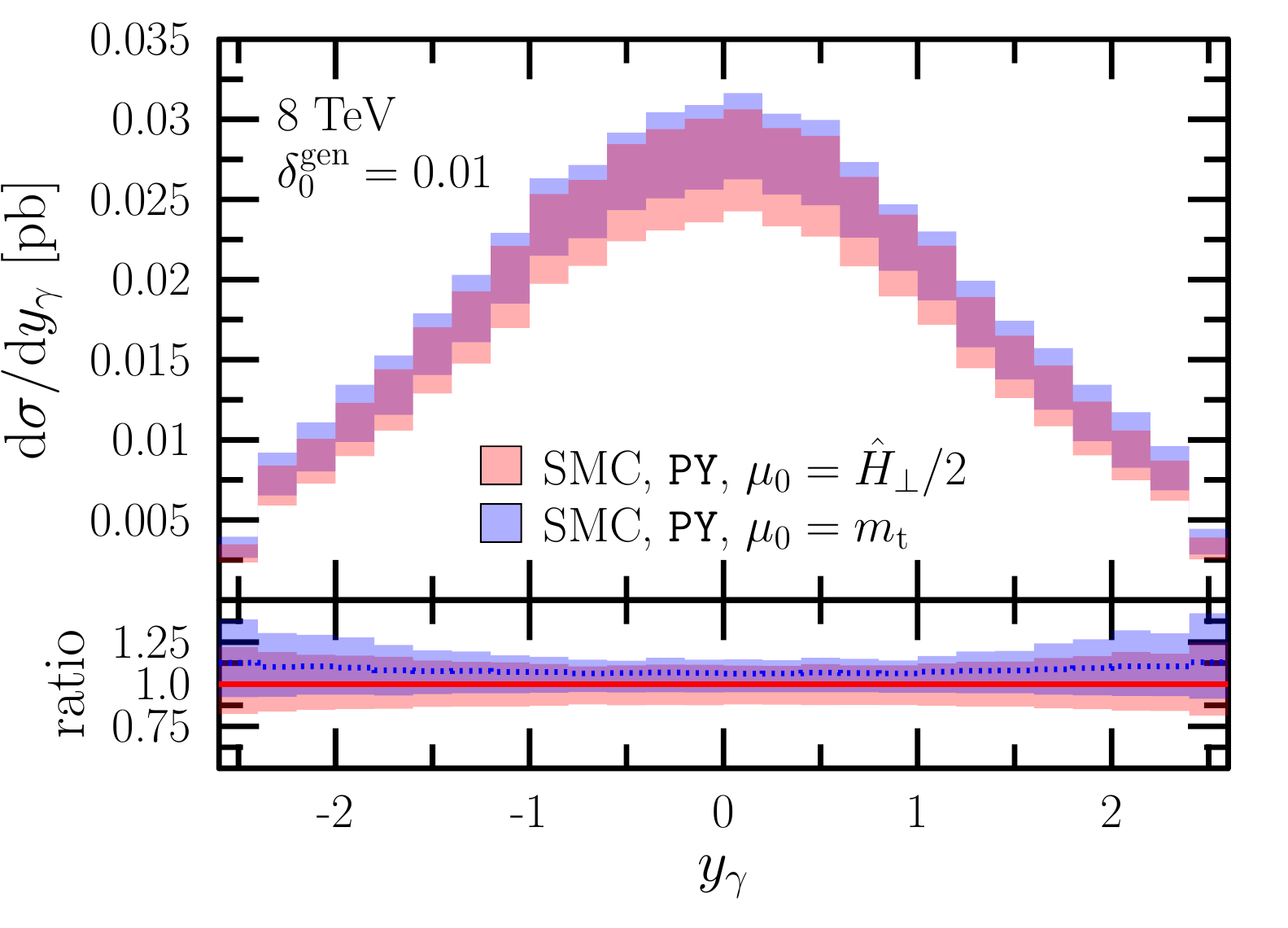}
\includegraphics[width=0.50\textwidth]{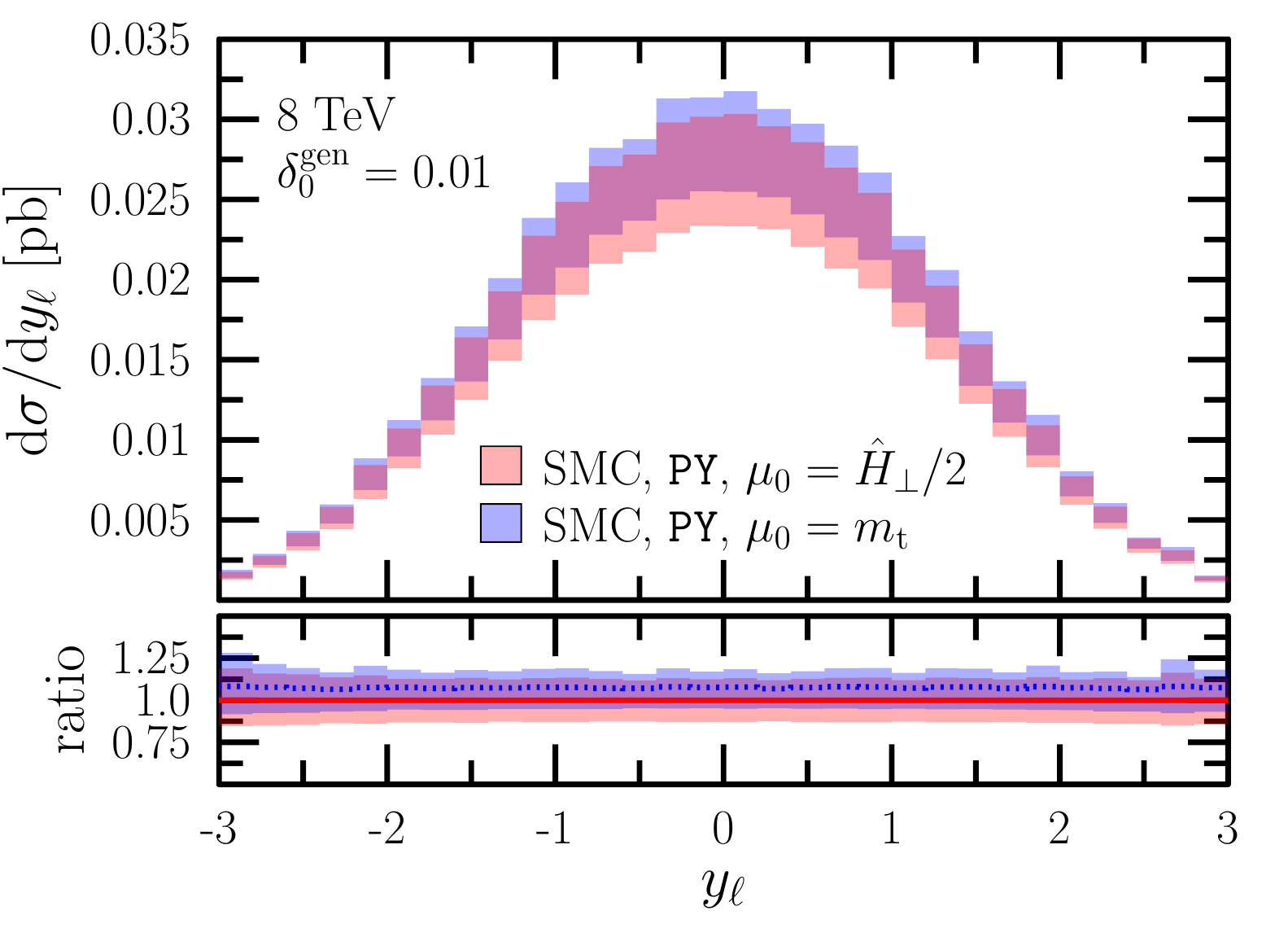}
\caption{\label{fig:SMCpred-y-gamma-lept} The same as 
\fig{fig:SMCpred-pt-gamma-tq} but for the rapidity distributions of
the photon and (anti)lepton.}
\end{figure}

If we turn our attention to the separations between the photon and the
t-quark, as well as between the photon and the charged lepton, measured
in the rapidity--azimuthal angle plane, we do not find significant
difference between the two scale choices, as seen in
\fig{fig:SMCpred-dR-gamma-tq-lept}. The static scale gives somewhat
higher cross section and a slightly narrower uncertainty band below
$\Delta R = \pi$ and larger scale dependence above.
Similar conclusions can be drawn from the  rapidity distributions 
for the photon and the (anti)lepton shown in \fig{fig:SMCpred-y-gamma-lept}.
In general, the scale dependence is moderate, below 20\,\% for both scale
choices and all observables, except for the predictions at large
transverse momenta with the static scale.

\section{Conclusions}

In this paper we presented a new way to make predictions for the
hadroproduction of isolated photons which uses event samples that
emerge in simulations aimed at matching predictions at NLO accuracy
with PS. Our approach uses only the direct-photon contribution, i.e.~we
neglect the fragmentation. We demonstrated that the presence of a
sufficiently small smooth isolation of the direct photons, applied
during generation of the events, does not affect the physical
predictions, within the numerical accuracy of the calculation. Hence it
can be used to generate sufficiently inclusive pre-showered event
samples. The pre-showered events obtained this way can be further
showered and hadronized to obtain differential distributions at the
hadronic stage, which include NLO QCD corrections in the hard process,
and either smooth or standard experimental photon isolation can be
applied.

Using the \powheg\ method one can make predictions at various
stages of the event simulation. In particular, for most of the
phenomenologically interesting distributions we estimate fairly small
(about 10\,\%, or less) corrections for the \ttbar$\gamma$ final state
due to the parton shower. We also studied the dependence of our
predictions on the renormalization and factorization scales and found
small and rather uniform scale dependence for the default scale 
$\hat{H}_\perp/2$.

The events generated with a loose generation isolation contain only
perturbative information, we neglected the non-perturbative
(fragmentation) contribution. We argued that making the generation 
isolation looser, the fragmentation contribution should decrease. 
Therefore in case of sufficiently loose generation isolation the
fragmentation can be neglected within the expected uncertainty of matched
NLO+PS predictions if the photon is harder than the accompanying jets.
This statement is trivially true if the experimental isolation is a
tighter version of the smooth isolation than that employed for event
generation.  We demonstrated the fragmentation also becomes negligible
in the case of hadroproduction of a $W$ boson in association with a
hard isolated photon by comparing our predictions to measured data if
the photon is harder than the accompanying jets.  Our method is
completely general and can be used to any process with isolated hard
photons in the final state, in particular also for \ttbar~production in
association with hard isolated photons.  

\section*{Acknowledgments}
This research was supported by
the Hungarian Scientific Research Fund grant K-101482,
the European Union and the European Social Fund through
Supercomputer, the national virtual lab TAMOP-4.2.2.C-11/1/KONV-2012-0010
and the
LHCPhenoNet network PITN-GA-2010-264564
projects.


\providecommand{\href}[2]{#2}\begingroup\raggedright\endgroup
\end{document}